\documentclass[usenatbib]{mnras}
\usepackage{graphicx,amsmath,,amssymb}
\usepackage{xspace}
\usepackage[english]{babel}

\def \etal {et~al.~}

\newcommand{\hMpc}{{\ifmmode{h^{-1}{\rm Mpc}}\else{$h^{-1}$Mpc}\fi}}
\newcommand{\hkpc}{{\ifmmode{h^{-1}{\rm kpc}}\else{$h^{-1}$kpc}\fi}}
\newcommand{\kpc}{{\ifmmode{ {\rm kpc} }\else{{\rm kpc}}\fi}}
\newcommand{\kms}{{\ifmmode{ {\rm km\,s^{-1}} }\else{ ${\rm km\,s^{-1}}$ }\fi}}
\newcommand{\hMsun}{{\ifmmode{h^{-1}{\rm {M_{\odot}}}}\else{$h^{-1}{\rm{M_{\odot}}}$}\fi}}
\newcommand{\Msun}{{\ifmmode{{\rm M}_{\odot}}\else{${\rm M}_{\odot}$}\fi}}
\newcommand{\Mhalo}{{\ifmmode{M_{\rm halo}}\else{$M_{\rm halo}$}\fi}}
\newcommand{\Rvir}{{\ifmmode{R_{\rm vir}}\else{$R_{\rm vir}$}\fi}}
\newcommand{\Mstar}{{\ifmmode{M_{\rm star}}\else{$M_{\rm star}$}\fi}}
\newcommand{\Vrot}{{\ifmmode{V_{\rm rot}}\else{$V_{\rm rot}$}\fi}}
\newcommand{\ltsima}{$\; \buildrel < \over \sim \;$}
\newcommand{\gtsima}{$\; \buildrel > \over \sim \;$}
\newcommand{\lsim}{\lower.5ex\hbox{\ltsima}}
\newcommand{\gsim}{\lower.5ex\hbox{\gtsima}}

\def\beqa{\begin{eqnarray}}
\def\eeqa{\end{eqnarray}}
\def\LCDM{\ensuremath{\Lambda}CDM}

\def\head{ \vbox to 0pt{\vss \hbox to 0pt{\hskip 440pt\rm
      LA-UR-10-07069\hss} \vskip 25pt}}

\def \kms {\ifmmode  \,\rm km\,s^{-1} \else $\,\rm km\,s^{-1}  $ \fi }
\def \kpc {\ifmmode  {\rm kpc}  \else ${\rm  kpc}$ \fi  }  
\def \hkpc {\ifmmode  {h^{-1}\rm kpc}  \else ${h^{-1}\rm kpc}$ \fi  }  
\def \hMpc {\ifmmode  {h^{-1}\rm Mpc}  \else ${h^{-1}\rm Mpc}$ \fi  }  
\def \Mpch {\ifmmode  {h^{-1}\rm Mpc}  \else ${h^{-1}\rm Mpc}$ \fi  }  
\def \Msun {\ifmmode {\rm M}_{\odot} \else ${\rm M}_{\odot}$ \fi} 
\def \hMsun {\ifmmode h^{-1}\,\rm M_{\odot} \else $h^{-1}\,\rm M_{\odot}$ \fi}

\def \LCDM {\ifmmode \Lambda{\rm CDM} \else $\Lambda{\rm CDM}$ \fi}
\def \sig8 {\ifmmode \sigma_8 \else $\sigma_8$ \fi} 
\def \OmegaM {\ifmmode \Omega_{\rm m} \else $\Omega_{\rm m}$ \fi} 
\def \Omegab {\ifmmode \Omega_{\rm b} \else $\Omega_{\rm b}$ \fi} 
\def \OmegaL {\ifmmode \Omega_{\rm \Lambda} \else $\Omega_{\rm \Lambda}$\fi} 
\def \Deltavir {\ifmmode \Delta_{\rm vir} \else $\Delta_{\rm vir}$ \fi}
\def \rhocrit {\ifmmode \rho_{\rm crit} \else $\rho_{\rm crit}$ \fi}
\def \rhou {\ifmmode \rho_{\rm u} \else $\rho_{\rm u}$ \fi}
\def \zc {\ifmmode z_{\rm c} \else $z_{\rm c}$ \fi}

\def\lcdm{\ensuremath{\Lambda\textrm{CDM}}\xspace}  
\def\omegam{\ensuremath{\Omega_\textrm{m}}\xspace}

\def\aj{AJ}%
\def\apj{ApJ}%
\def\apjl{ApJ}%
\def\apjs{ApJS}%
\def\ao{Appl.~Opt.}%
\def\aap{A\&A}%
\def\jcap{J. Cosmology Astropart. Phys.}%
\def\mnras{MNRAS}%
\def\nat{Nature}%
\def\head{ .ps \vbox to 0pt{\vss \hbox to 0pt{\hskip 440pt\rm
      LA-UR-10-07069\hss} \vskip 25pt}} 

\def \spose#1{\hbox  to 0pt{#1\hss}}  
\def \lta{\mathrel{\spose{\lower 3pt\hbox{$\sim$}}\raise 2.0pt\hbox{$<$}}}
\def \gta{\mathrel{\spose{\lower 3pt\hbox{$\sim$}}\raise 2.0pt\hbox{$>$}}}

\title[Andromeda analog thin Planes of Satellites]
{Simulated \lcdm analogues of the thin Plane of Satellites around the Andromeda galaxy are not kinematically coherent structures}

\author[T. Buck \etal] {Tobias Buck$^{1}$\thanks{E-mail:
    buck@mpia.de}\thanks{Member of the International Max Planck Research School for Astronomy and Cosmic Physics at the University of Heidelberg, IMPRS-HD, Germany.},
    Aaron A. Dutton$^{2}$\thanks{E-mail:
    dutton@nyu.edu}, Andrea V. Macci\`o$^{1,2}$\thanks{E-mail:
    maccio@nyu.edu}
    \\
$^1$Max-Planck-Institut f\"ur Astronomie, K\"onigstuhl 17, 69117 Heidelberg, Germany\\
$^2$New York University Abu Dhabi, PO Box 129188, Abu Dhabi, UAE
}

\begin{document}

\date{Accepted 2016 May 19. Received 2016 May 19; in original form 2016 February 21}

\pagerange{\pageref{firstpage}--\pageref{lastpage}} \pubyear{2016}

\maketitle

\label{firstpage}


\begin{abstract}
A large fraction of the dwarf satellites orbiting the Andromeda galaxy are surprisingly aligned in a thin, extended and apparently kinematically coherent planar structure. Such a structure is not easily found in simulations based on the Cold Dark Matter model (\lcdm). Using 21 high resolution cosmological simulations we analyse the kinematics of planes of satellites similar to the one around Andromeda. We find good agreement when co-rotation is characterized by the line-of-sight velocity. At the same time, when co-rotation is inferred by the angular momenta of the satellites, the planes are in agreement with the plane around our Galaxy. We find such planes to be common in our high concentration haloes. The number of co-rotating satellites obtained from the sign of the line-of-sight velocity shows large variations depending on the viewing angle and is consistent with that obtained from a sample with random velocities. We find that the clustering of angular momentum vectors of the satellites in the plane is a better measure of the kinematic coherence. Thus we conclude that the line-of-sight velocity is not well suited as a proxy for the kinematical coherence of the plane. Analysis of the kinematics of our planes shows a fraction of $\sim$30\% chance aligned satellites. Tracking the satellites in the plane back in time reveals that these planes are a transient feature and not kinematically coherent as would appear at first sight. Thus we expect some of the satellites in the plane around Andromeda to have high velocities perpendicular to the plane.
\end{abstract}

\noindent
\begin{keywords}

  cosmology: dark matter galaxies: individual Andromeda - dwarf - kinematics and dynamics - formation methods:N-body simulation

 \end{keywords}

\section{Introduction} \label{sec:introduction}

\begin{table}
\label{tab:sims}
\begin{center}
\caption{Properties of our 21 high resolution simulations.}
\begin{tabular}{p{0.04\linewidth} p{0.125\linewidth} p{0.15\linewidth} p{0.075\linewidth} p{0.075\linewidth} p{0.04\linewidth} p{0.16\linewidth}}
		\hline\hline
		Halo & Box size & $M_{200}$ & $R_{200}$ & $c_{\rm 200}$ & $N_{\rm 200}$ & Force soft.\\
		 & [Mpc/$h$] & [$10^{12} M_{\odot}/h$] & [kpc/$h$] & & [$10^6$] & [kpc/$h$]\\
		\hline
		(A)& 30 &0.736& 146.82 & 13.62 &8.3 & 0.25 \\
		(B) & 80 &1.003& 162.78 &12.03 &7.6 & 0.29 \\
		(C)& 30 &1.321& 178.44 & 11.29 &10.2 & 0.25 \\
		(D)& 60 &0.890& 156.45 & 7.65 &7.3 & 0.25 \\
		(E)& 60 &0.570& 134.85 & 8.19 &6.4 & 0.27 \\
		(F)& 60 &0.831& 152.88 & 8.69 &9.4 & 0.25 \\
		(G)& 60 &0.637& 139.93 & 4.18 &7.2 & 0.25 \\
		(H)& 60 &0.813& 151.78 & 7.19 &9.2 & 0.25 \\
		(I)& 60 &0.535& 132.02 &17.19 &6.01 & 0.25 \\
		(J)& 60 &0.907& 157.41 & 6.93 &7.5 & 0.27 \\
		(K)& 60 &0.654& 141.20 & 5.63 &7.4 & 0.25 \\
		(L)& 60 &0.872& 155.37 &11.63 &7.2 & 0.28 \\
		(M)& 60 &0.661& 141.65 & 7.65 &7.5 & 0.25 \\
		(N)& 60 &0.626& 139.12 &13.30 &7.1 & 0.25 \\
		(O)& 60 &0.760& 148.44 & 9.14 &8.6 & 0.25 \\
		(P)& 60 &0.864& 154.90 &13.93 &9.8 & 0.25 \\
		(Q)& 60 &0.702& 144.54 &10.67 &7.9 & 0.25 \\
		(R)& 80 &1.389& 181.47 & 11.35 &5.4 & 0.36 \\ 
		(S)& 45 &0.848& 153.93 & 6.86 &6.4 & 0.29 \\	
		(T)& 30 &1.028& 164.12 & 6.37 &11.6 & 0.25 \\
		(U)& 30 &0.922& 158.28 & 7.82 &10.4 & 0.25 \\	
		\hline
\end{tabular}
\end{center}
\end{table}

The currently favored model of structure formation, the Cold Dark Matter model together with Dark Energy $\Lambda$ (\lcdm), is in outstanding accordance with observations on large scales \citep{Tegmark2004,Springel2005} and describes the collapse of dark matter haloes, which are the hosts of galaxies. However, its constraints on the structure of baryons on galactic and sub-galactic scales are considerably less evidential. In the past two decades there have been several problems reported on these scales, e.g. the ``missing satellite problem" \citep{Klypin1999,Moore1999a}, the ``cusp-core problem"
\citep{Moore1994,DeBlok2001} or the ``too big to fail problem" \citep{Boylan2011}. A potential solution to these issues is the complex interplay of baryonic physics and Dark Matter on small scales. Cosmological simulations
including baryons (gas and stars) have been able to ease those problems and to bring the CDM model in agreement with observations \citep{Benson2007,Maccio2010a,Governato2010,DiCintio2014,Brooks2014,Chan2015,Tollet2016,Dutton2016}.
But still not all observed galactic structures can be explained by the \lcdm model. 
Despite explaining perfectly the anisotropic nature of the Cosmic Web on larger scales the anisotropic dwarf galaxy distribution lacks conclusive explanation.

Since \citet{LyndenBell1976} found the first indication for a planar alignment of  the classical dwarfs and stellar streams of our Milky Way (MW) much progress in discovering further satellites and more planes of satellites has been made. Proper motion measurements showed that many members of the MW plane share a common rotation direction and thus it has been concluded that this plane is rotationally stabilized \citep{Metz2008,Pawlowski2013}. Similarly recent observations of dwarf satellite galaxies around Andromeda (M31) revealed an anisotropic spatial distribution of the satellites around our neighboring galaxy \citep{Koch2006,McConnachie2006,Conn2013}. These observations found roughly half of the satellites of M31 discovered in the PAndAS survey \citep{PANDAS} (15 out of 27) to align in a thin plane of thickness ($12.6 \pm0.6$) kpc with a projected diameter of $\sim$280 kpc viewed nearly edge-on from Earth. Furthermore the analysis of the line-of-sight velocities of the satellites in the plane conducted by \citet{Ibata2013} showed that out of the 15 members in the plane 13 seem to share the same sense of rotation. Besides the plane around Andromeda there might be further observational indication for similar planes around other galaxies like Centaurus A \citep{Tully2015} or M81 \citep{M812013} as well as NGC 3109 \citep{NGC31092013} albeit much less convincing.

Many works have focussed on an explanation of the anisotropic distribution of satellites in \lcdm \citep{Libeskind2005,Zentner2005,Kang2005,Libeskind2009,Libeskind2011,Deason2011,Hammer2013,Wang2013,Smith2016} and found that this seems to be a common prediction. Flattened satellite distributions can arise from the filamentary accretion of substructure onto the main halo \citep{Aubert2004,Libeskind2005} and the spatial information of the anisotropic accretion is retained in the distribution of the satellites in phase space \citep{Libeskind2009,Lovell2011,Libeskind2011}. Using the fact, that early forming haloes are more likely to reside at the nodes of intersections of dark matter filaments \citep{Dekel2009}, \citet{Buck2015} (in the following called Paper I) could show that these haloes have a higher incidence of planes resembling the one around Andromeda with satellites in the plane being accreted onto the main halo via filaments from opposite directions.

Although anisotropies are predicted by \lcdm due to the nature of the Cosmic Web and the presence of filaments, apparently rotating satellite distributions resembling the one around MW and Andromeda are not easily found in \lcdm.
An investigation by \citet{Wang2013} of the Aquarius and Millenium II simulations for satellite distributions consisting of 11 satellites similar to the one around our Milky Way shows that 5-10\% of the satellite distributions in their sample are as flat as the one around our Milky Way. When demanding the same rotational coherence as for the Milky Way satellites, with 8 out of 11 co-orbiting \citep{Pawlowski2013}, and using as well the Millennium II simulation, \citet{Pawlowski2014a} found that only $\sim$0.1\% of the satellite systems show similar planes. The same result is found when searching for planes similar to the one around Andromeda, consisting of 15 satellites in the plane with 13 co-rotating ones. Again using the Millennium II simulation, \citet{Ibata2014} and \citet{Pawlowski2014a} found that only $\sim$0.1\% of all \lcdm satellite systems show similar distributions as the one around Andromeda. Recent work by \citet{Gillet2015} using hydrodynamical simulations of Local Group analogues also found flattened satellite systems but not as rich in number as the one around Andromeda. However, a work by \citet{Cautun2015} pointed out the large ambiguity in defining planes of satellites in terms of thickness vs. number of satellites in the plane. That work shows that, when accounting for this problem, $\sim$10\% of all \lcdm haloes host more prominent planes compared to the Local Group. 

While all of these studies focussed on either finding planes of satellites most similar to the ones observed, explaining the mechanism leading to planes of satellites or trying to quantify their occurrence, less work has been spend on investigating the actual kinematics of the planes produce by \lcdm. The observations of the plane around Andromeda are so striking due to the high kinematical coherence of the satellites in the plane. The high degree of co-rotation inferred from the line-of-sight velocities makes the plane a challenge for \lcdm. But, for the conclusion of kinematical coherence via the line-of-sight velocity, only one out of three velocity components is used. Therefore in this work we analyse in detail the kinematics of the satellite planes found in high-resolution cosmological ``zoom-in" Dark Matter only simulations of the \lcdm model. The planes used for this work resemble the ones observed in the Local Group and were previously used in Paper I to show the filamentary accretion of satellites in the plane of early forming haloes. 

This paper is organized as follows: In \S2 we present the simulations,
including the host halo selection criteria and our plane finding
algorithm and show some visual impressions of planes. In section \S3 we then present the results of our spatial analysis, such as the thickness of the plane vs. its radial extent and the number of satellites found in the plane. In \S4 we then move on to a detailed kinematical analysis of the planes with a comparison between the kinematical coherence inferred from line-of-sight velocities and from angular momentum counting. Furthermore we investigate the time evolution of the planes and finally in \S5 we summarize and discuss our findings and present our conclusions in \S6.

\section{Simulations} \label{sec:simulation}

For this work we used the suite of 21 high resolution ``zoom-in'' Dark Matter only
simulations from Paper I. These simulations include Andromeda-mass haloes in the mass range ($7.4\times10^{11} <M_{200}/[M_{\odot}] < 2.2\times10^{12}$), where the halo mass was defined with respect to 200 times the critical density of the universe. These values are in agreement with recent measurements of the mass of Andromeda from dynamical modeling the Local Group which derives a mass of $1.24^{+1.1}_{-0.65}$ $M_{\odot}$ \citep{Penarrubia2014} for Andromeda and its dark matter halo.
The Dark Matter haloes re-simulated for that work were selected from four cosmological boxes of side length 30, 45, 60 and 80 $h^{-1}$Mpc from \citet{Dutton2014}, who used
cosmological parameters from the Planck Collaboration
\citeyearpar{Planck}: $\omegam=0.3175$, $h=0.671$, $\sigma_8=0.8344$,
$n=0.9624$.  The simulations were then evolved to redshift $z=0$ with the N-body code {\sc{pkdgrav2}} \citep{Stadel2001,Stadel2013}.
Initial conditions for the ``zoom-in" simulations were created using
a modified version of the {\sc{grafic2}} package \citep{Bertschinger2001} as described in
\citep{Penzo2014}. The refinement level was chosen to maintain a roughly
constant relative resolution, e.g. $\sim 10^7$ dark matter particles
per halo with particle masses of $\sim 10^5 h^{-1}$M$_{\odot}$. This
allows us to reliably resolve substructure down to $\sim 10^7
h^{-1}$M$_{\odot}$, comparable to the expected dynamical masses of dwarf satellites in the local group \citep[Fig. 10]{McConnachie2012}.  The force softening of the simulations ranges
between $0.25$ $h^{-1}$ kpc and $0.36$ $h^{-1}$ kpc (see Table: \ref{tab:sims}).

\begin{figure}
\includegraphics[width=.49\textwidth]{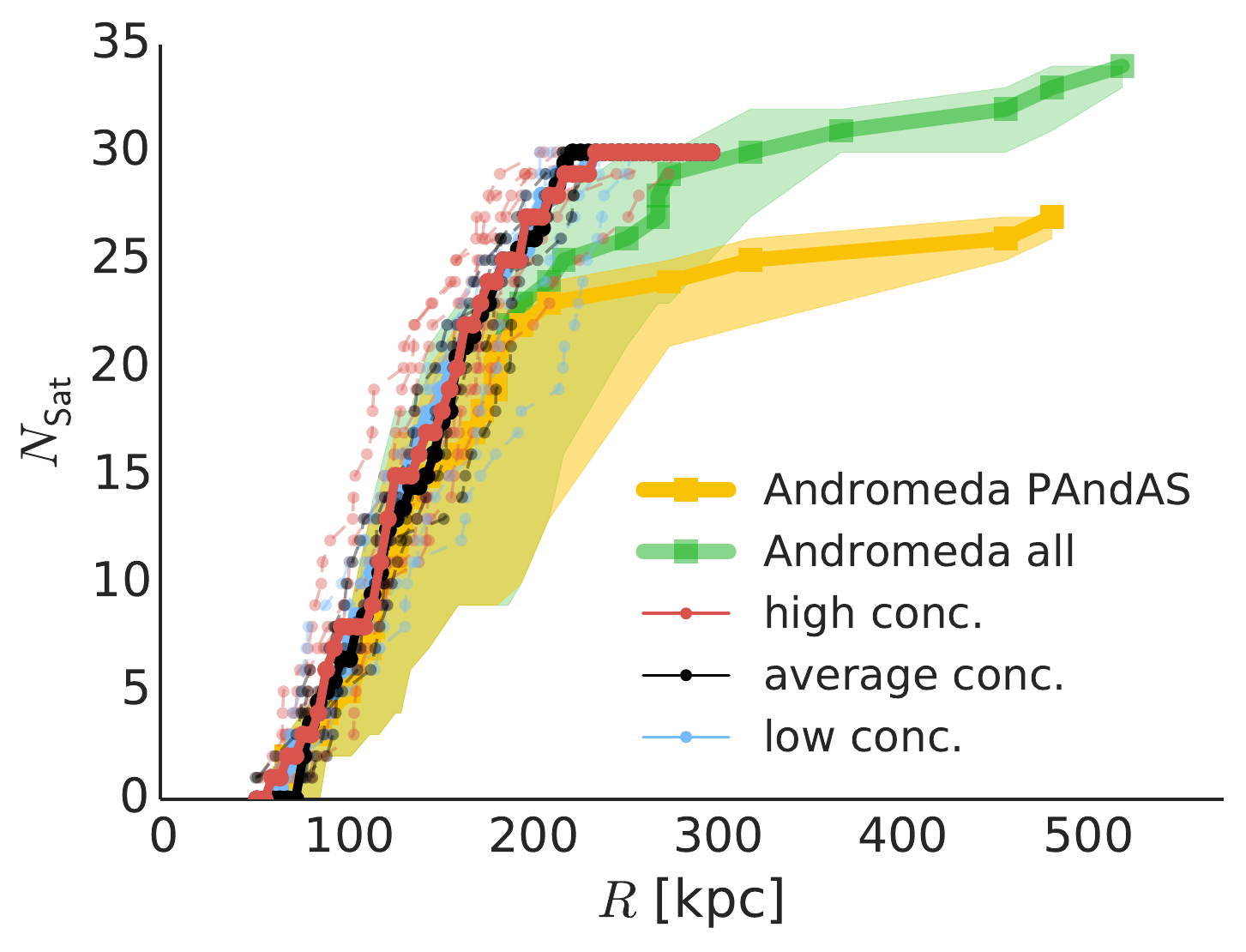}
\vspace{-.5cm}
\caption{The present day cumulative radial distribution of the 30 most massive satellites at the infall time. The red line shows the median of the high concentration sample, the black line shows the median of the average concentration sample and the blue line shows the same for the low concentration sample. The whole sample is shown as light colored lines. The yellow line indicates the cumulative radial distribution of the satellites around Andromeda within the PAndAS footprint. The shaded yellow area marks the measurement uncertainty of the radial distribution of the satellites around Andromeda by using the maximal and minimal radius of every satellite given by their measurement uncertainties. The green line together with the green shaded area shows the same but including several known satellites outside the PAndAS survey area.}
\label{fig:rad_dist}
\end{figure}

\subsection{Halo concentration and formation time}
As established in Paper I, aside from halo mass, the only other selection criteria for the
haloes was the concentration, which was used as a proxy for halo formation time
\citep{Wechsler2002}. The reasoning behind this correlation is, that at a fixed mass at the
present time, early forming haloes are more likely to form at the
nodes of intersections of a few filaments of the cosmic web, while
typical haloes tend to reside inside such filaments \citep{Dekel2009}. Therefore one
then expects that, rare, early forming haloes would accrete their
satellites from a few streams that are narrow compared to the halo
size (see Paper I, Figure 5), while later forming and thus more typical haloes accrete satellites from a wider angle in a
less anisotropic manner (see also \cite{Libeskind2014} for redshift dependence of anisotropic accretion).\par

A roughly equal number of high, average and low concentration haloes was selected, where we aimed at sampling the whole range of concentrations possible. Especially haloes with concentrations far offset from the main relation were selected (see left panel of Figure 1 of Paper I). Our high concentration haloes have on average an offset of
about $2\sigma$ from the mean relation. This means these haloes are
the rarest 2.3\% of the whole population. For a random sampling of
haloes, it would thus require $\sim 40$ simulations to recover one such
rare halo. This helps us to explain why previous high resolution
simulations were unable to reproduce the observed properties of the
satellite distribution around the Andromeda galaxy: they simply did
not sample enough haloes to find the rarer earliest forming ones.

\begin{figure*}
\centering
\includegraphics[width=\textwidth]{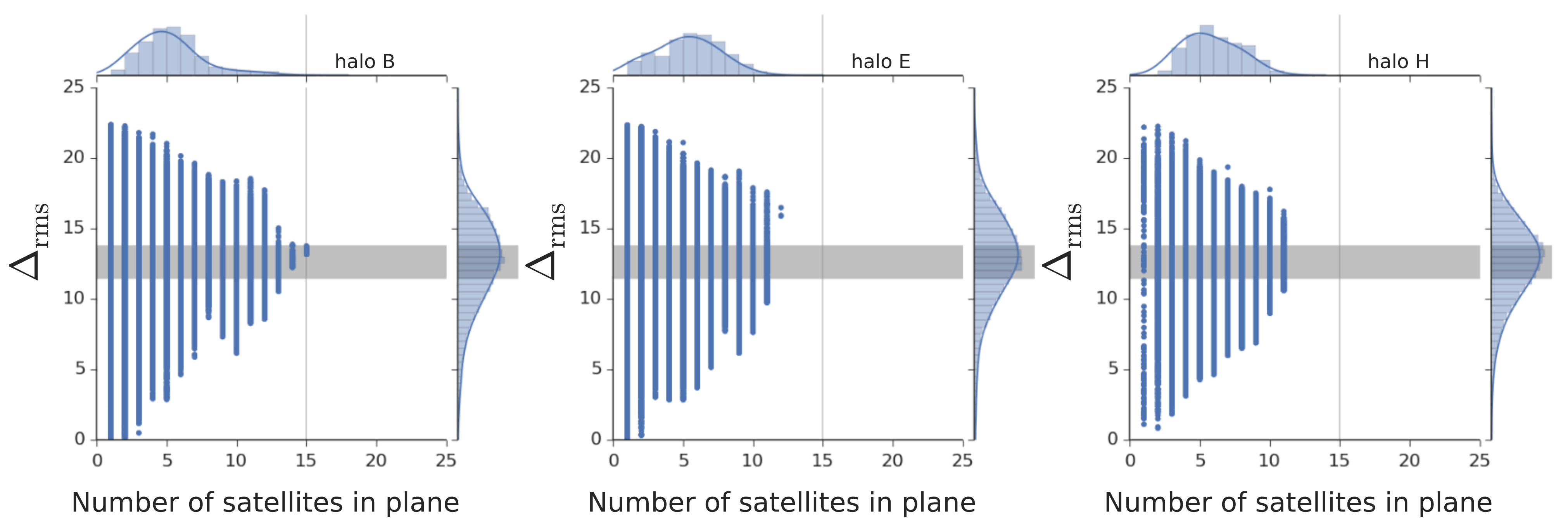}
\vspace*{-.5cm}
\caption{The distribution of plane parameters for a particular number of satellites in the plane $N_{\rm in}$ and the rms value, $\Delta_{\rm rms}$, of the associated plane. The \emph{top marginal histogram} shows the cumulation along the y-axis, and the \emph{right marginal histogram} shows the cumulation along the x-axis. The solid blue line in the marginal plots shows a kernel density estimation of the probability distribution function. The \emph{grey line} shows a value of 15 satellites in the plane and the grey shaded area shows the measurement uncertainty of $\Delta_{\rm rms}= 12.6 \pm 0.6$ kpc of the observed thickness of the plane around Andromeda. The \emph{left panel} shows a high concentration halo, the \emph{middle panel} an average concentration halo and the \emph{right panel} shows a low concentration halo.}
\label{fig:2d_significance}
\end{figure*}

\subsection{Satellite selection}

The substructure within the high-resolution region was identified using the \emph{Amiga Halo Finder}\footnote{http://popia.ft.uam.es/AHF/Download.html} 
\citep[AHF]{Knollmann2009}, which is capable of finding sub-haloes as well as dealing with different particle masses. \emph{AHF} identifies structures based on density estimates calculated with an adaptive refinement technique and performs an iterative process to remove unbound particles within the virial radius, defined as the radius where the average density equals 200 times the critical density $\bar{\rho}(r_{\rm vir})=200\rho_{\rm crit}$. The \emph{Amiga Halo Finder} identifies hundreds of resolved sub-haloes in the high-resolution simulations from which the actual luminous galaxies have to be chosen. Galaxy formation models robustly predict the luminous sub-haloes to be the ones most massive at infall times \citep{Kravtsov2004,Conroy2006,Vale2006}. Thus, like in Paper I, we selected a sample of the 30 most massive sub-haloes at the time of the accretion via the maximum circular velocity $v_{\rm max}$. To find the maximum circular velocity at the time of accretion for every satellite in the simulations we use a fitting formula from \citet[see their formula A.2]{Anderhalden2013}. With this formula, the maximum circular velocity at the time of accretion can be estimated from the present day value and the orbital energy of the satellite. 

For this analysis we decided to choose a more physical volume instead of modeling the PAndAS \citep{PANDAS} footprint, and thus restrict ourselves to sub-haloes within the virial radius of the host halo ($\sim 250$ kpc) at the present time. Following the observations, we exclude satellites within the innermost 30 kpc. Selection of the 30 most massive satellites at the infall time results in present day satellite masses in the range ($4.4\times10^{7} < M_{200}/[M_{\odot}] < 1.5\times10^{10}$). We select a number of 30 satellites instead of 27 like it is the case for the plane around Andromeda for two reasons: First, the number of 27 satellites around Andromeda results from the number of Dwarf galaxies found only in the PAndAS footprint but there are 9 additional satellites of Andromeda known \cite[see their Fig. 1]{Ibata2013}. Second, since the analysis of \citet{Ibata2013}, there have been 3 more satellites found around Andromeda with the Pan-STARRS1 3$\pi$ survey \citep{Martin2014}. So, to model the right number of satellites around Andromeda we would need to select either 36 or even 39 satellites, but to stay consistent with the amount of satellites used to find the plane we chose 30. We also identified planes for a number of 27 satellites and found that neither the thickness nor the number of satellites in the plane changes significantly.\par

Selecting host haloes of different concentrations might have an effect on the properties of the associated satellites. Namely leading to a more concentrated, more compact radial distribution of satellites in high concentration haloes compared to the ones in low concentration haloes. However, this is not the case for our satellite samples. There is no indication that high concentration haloes have on average a different radial distribution, compared to low and average concentration haloes or the observed satellites of Andromeda. In fact, the satellites of the haloes in the different samples show the same mean radius of $\sim130$kpc.
Figure \ref{fig:rad_dist} shows the present day cumulative radial distribution of the 21 satellite samples together with the observations plotted in yellow and green \citep{Conn2012,McConnachie2012}. The red, black and blue line shows the median radial distribution of the three sub samples of high, average and low concentration haloes. This Figure reveals that the selection of the 30 most massive satellites at infall time results in satellite samples occupying radii between (75$ \lesssim$ $R$/[kpc] $\lesssim$ 250) in agreement with the observed radial distribution of the 27 satellites around Andromeda up to a radius of $\sim$200 kpc. Inclusion of two satellites at radii of about $r\sim400$kpc into the observational sample leads to the fact that the cumulative radial distribution extends far further out and deviates from the simulated distribution at large radii. Comparison to the green line which includes known satellites around Andromeda not falling within the PAndAS footprint (IC10,A7,A28,A29,A6,IC1613,LGS3) shows that the discrepancy between observations and simulations might be due to the PAndAS footprint not covering the whole virial radius of Andromeda. 
A two sample Kolmogorov-Smirnov test on the median cumulative radial distribution of satellites of the different concentration samples shows with a p-value of 0.999 that the these samples are drawn from the same distribution. Thus there is no indication that high concentration haloes posses a more concentrated satellite distribution (see also Appendix A for more tests).

\subsection{Plane finding algorithm}

At $z=0$, we apply a plane finding algorithm (as described in \citet{Buck2015,Gillet2015,Ibata2014,Conn2013}) to the 30 most massive satellites at infall time, with the goal of identifying planes of satellites similar to the observed ones.
In order to find planes in the distribution of sub-haloes we generate
a random sample of planes defined by their normal vector. All planes
include the center of the main halo but we exclude satellites closer to the main halo center than 30 kpc. To uniformly cover the whole
volume we generate 100,000 random planes with a fixed thickness of
2$\Delta=30h^{-1}$ kpc. After specifying a plane we calculate the
distance of every satellite to this plane. A satellite is considered
to lie in the plane if its distance to the plane is smaller than
$\Delta$. For each plane we calculate the number of satellites in the
plane and its thickness $\Delta_{\rm rms}$ as the root-mean-square value of the satellites distances to the plane's mid-plane. We then
select for every number of satellites in the plane the one which is
thinnest and richest to analyse for kinematics.

The plane of satellites around Andromeda can be characterized by 4
parameters including the number of satellites in the plane ($N_{\rm in}$), the number of co-rotating satellites ($N_{\rm corot}$), the
thickness of the plane ($\Delta_{\rm rms}$, defined as the root-mean-square distance of satellites from the plane) and its extension on the plane of the sky ($\Delta_{\parallel}$). We define $\Delta_{\parallel}$ as the root-mean-square value of the distance of the satellites in the plane from the halo center when projected on the plane of the sky. For Andromeda these values are $N_{\rm in}=15$, $N_{\rm corot}=13$,
$\Delta_{\rm rms}= 12.6 \pm 0.6$ kpc \citep{Ibata2014} and $\Delta_{\parallel}=91.2^{+6.2}_{-11.7}$, calculated from the data given by \cite{Conn2012,McConnachie2012}.

\begin{figure}
\centering
\includegraphics[width=\columnwidth]{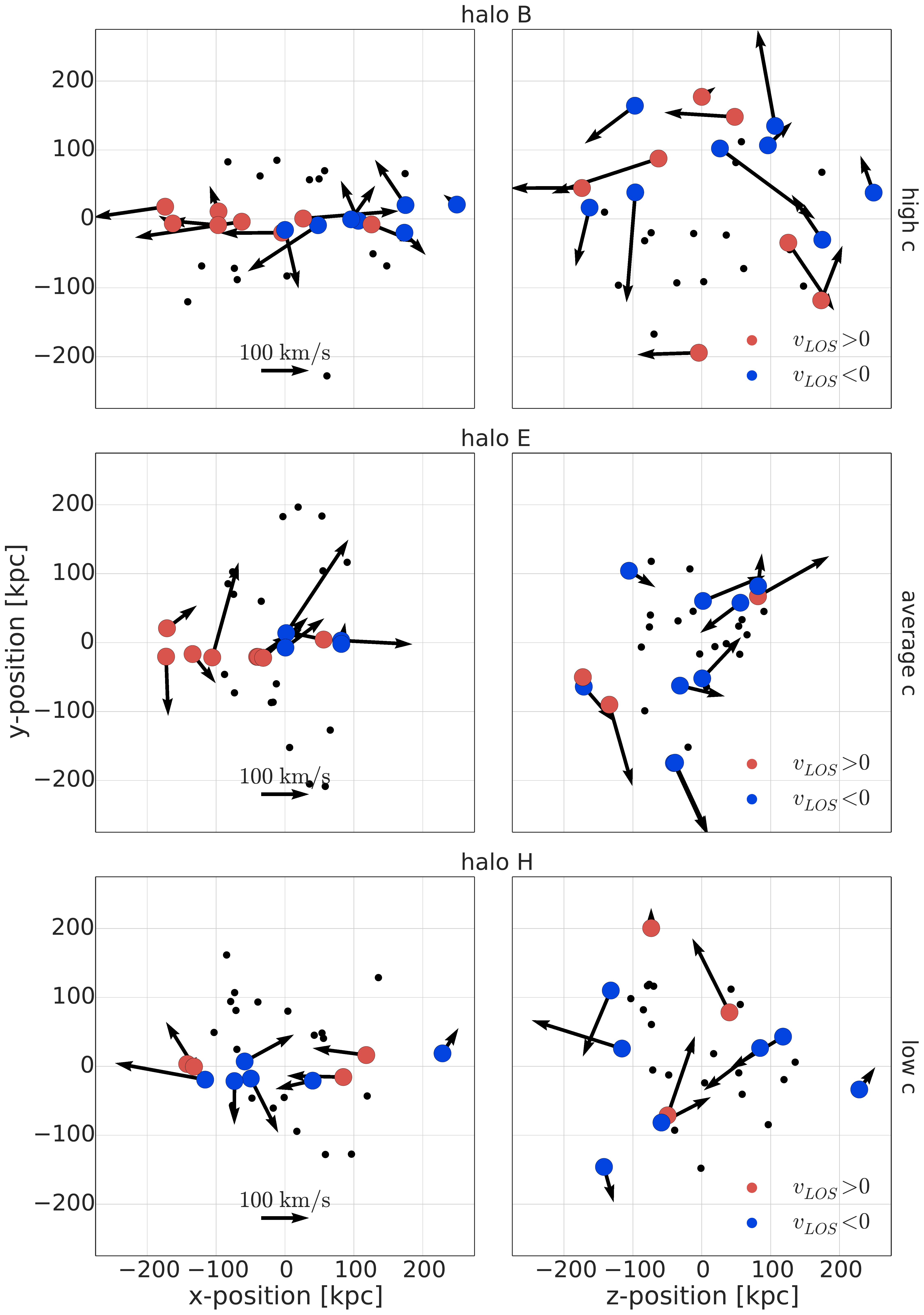}
\caption{A visual impression of the planes. The \emph{left panels} show the planes in edge-on view, while the \emph{right panels} show the planes in face-on view. The colored dots show satellites in the plane with the color indicating the line-of-sight velocity (blue approaching, red receding), where the line-of-sight is chosen to be perpendicular to the plane of the plot (left panels: along the z-axis, right panels: x-axis). Black dots show the satellites that lie outside the plane. \emph{Arrows} indicate the velocity perpendicular to the plane (left panels) and in the plane (right panels). The \emph{top row} shows a high concentration halo (halo B), the \emph{middle row} an average concentration halo (halo E) and the \emph{bottom row} shows a low concentration halo (halo H).}
\label{fig:planes}
\end{figure}

In the progress of the paper we will investigate the number of co-rotating satellites. This will be done for two different measures of co-rotation. Assuming the plane of satellites lies in the x-z plane (compare Figure \ref{fig:planes}), then when using the line-of-sight velocity as a proxy for co-rotation, like it is done in the observations, the number of co-rotating satellites is the maximum of counting all the satellites with x-position smaller than zero and line-of-sight velocity greater than zero and adding all the satellites with x-position greater than zero and line-of-sight velocity smaller than zero or doing the opposite. When co-rotation is inferred using the angular momentum vectors of the satellites the number of co-rotating satellites is the maximum of the number of satellites with positive z-component of the angular momentum vector or negative z-component.

\begin{figure*}
\begin{center}
\includegraphics[width=.49\textwidth]{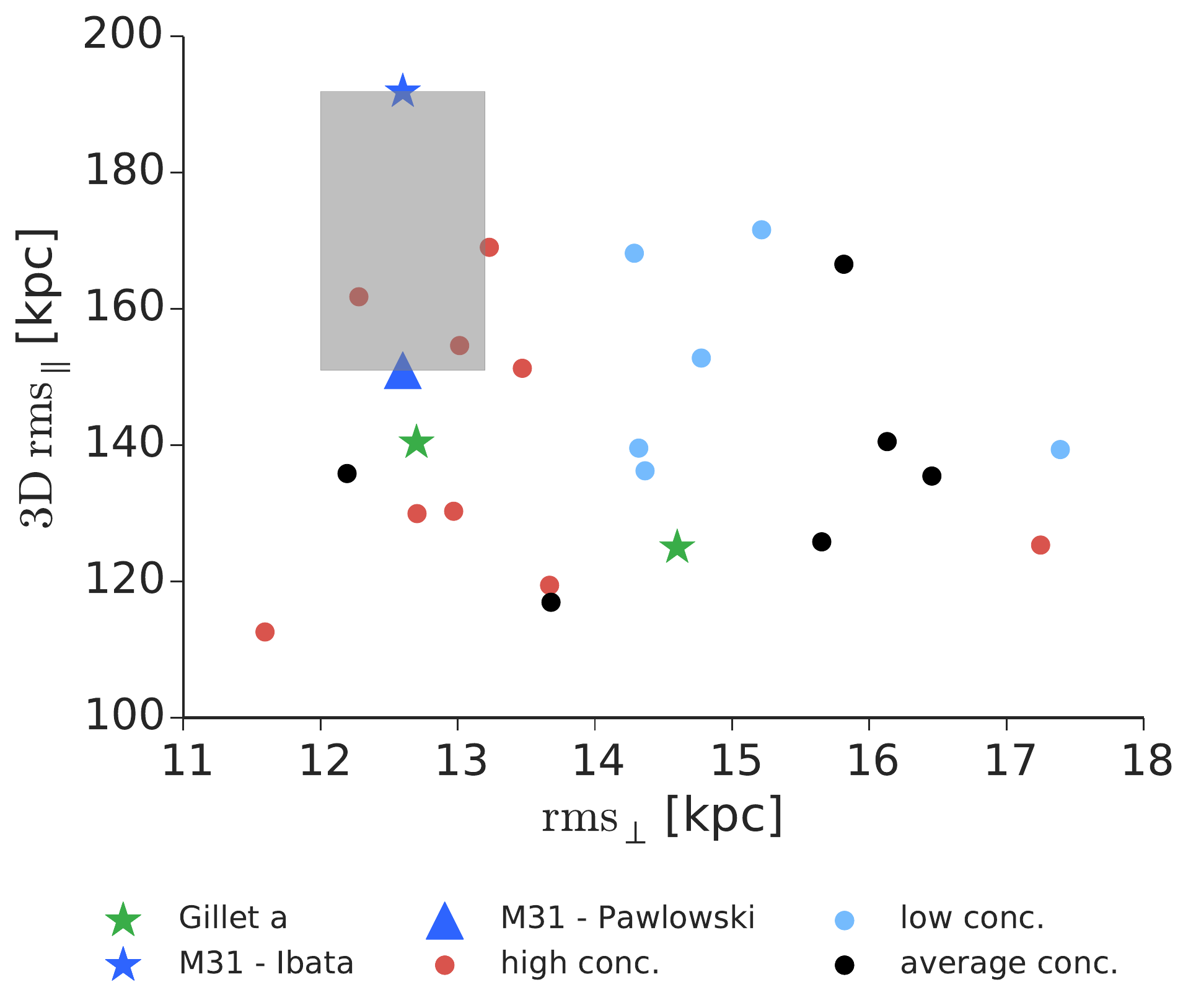}
\includegraphics[width=.49\textwidth]{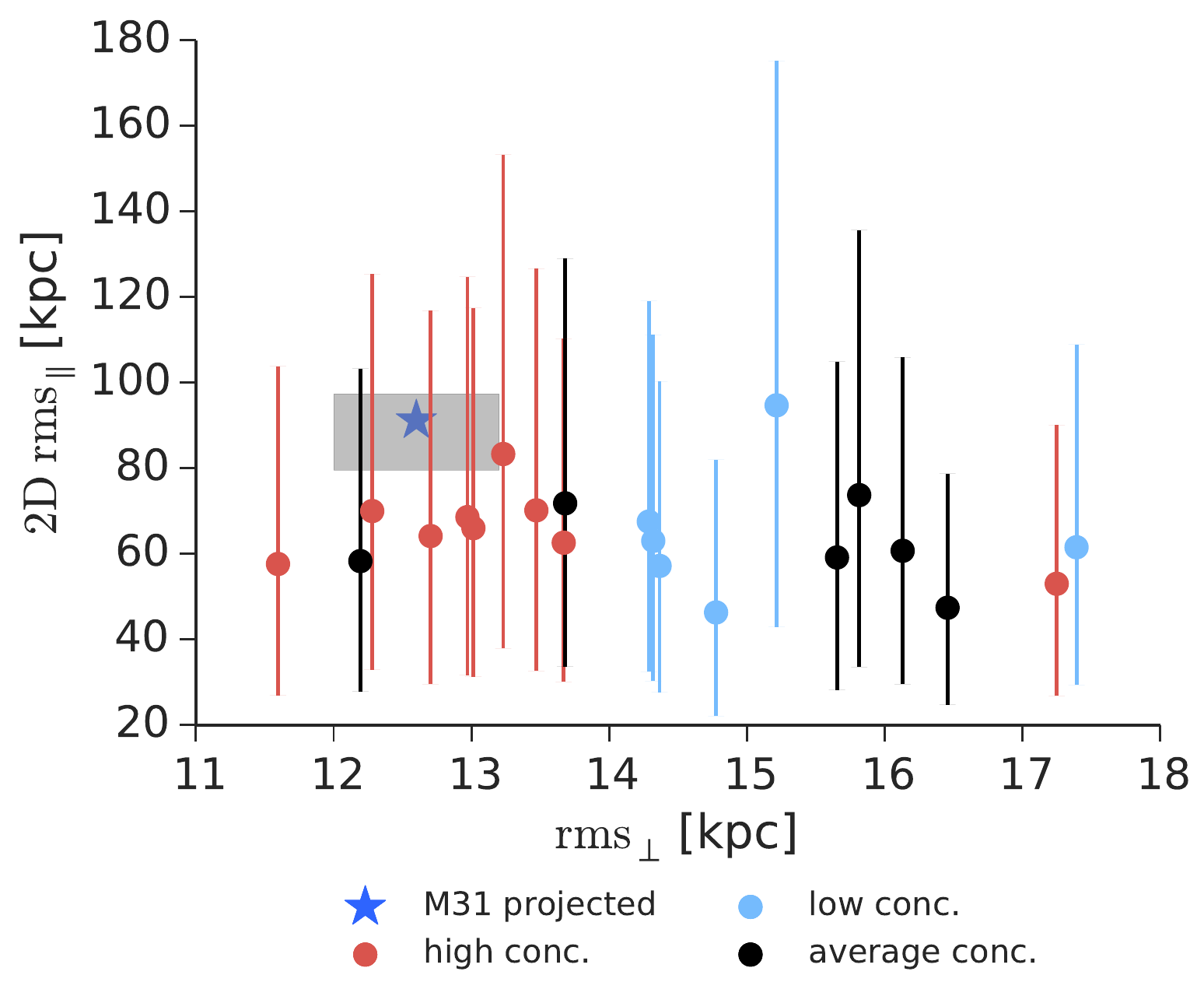}
\end{center}
\vspace{-.35cm}
\caption{\emph{Left panel}: The minimal root-mean-square thicknesses parallel and perpendicular to the richest planes found, color-coded by halo concentration. The values for Andromeda are indicated by the \emph{blue triangle} \citep{Pawlowski2014} and the \emph{blue star} \citep{Ibata2013}, with the grey shaded area indicating a nominal uncertainty of the perpendicular rms value of ($12.6\pm0.6$) kpc and the parallel rms calculated by the difference of the two values given in the literature. The values obtained by \citet{Gillet2015} for their two Andromeda analogues are plotted in green. \emph{Right panel}: The projected 2 dimensional root-mean-square thickness parallel and perpendicular to the richest plane found, color-coded by halo concentration. The value for Andromeda is indicated by the \emph{blue star} where the grey shaded area indicates the measurement uncertainties. For the perpendicular rms value this is given by ($12.6\pm0.6$) kpc and for the parallel one the value is $\Delta_{\parallel}=91.2^{+6.2}_{-11.7}$ kpc (this is the extension of the plane when projected on the plane of the sky), where the uncertainty is given by calculating the projected rms value with the upper and lower limit of the radial distance of each satellite from Andromeda. The error bars of the simulated satellite planes indicate the range of possible values if different viewing angles for the projection of the edge-on plane are chosen.}
\label{fig:rms_per_par}
\end{figure*}

\subsection{Distribution of plane parameters}

When applying our plane finding algorithm to the selected satellite samples of our simulations we obtain a variety of different planes characterized by different combinations of plane parameters. For example, it is possible to obtain different planes with the same number of satellites in them but with totally different thicknesses. This is shown in Figure \ref{fig:2d_significance} for three examples of our haloes. This figure shows a scatter plots of the plane parameters obtained. For every number of satellites in the plane, $N_{\rm in}$, the root-mean-square value for the plane thickness, $\Delta_{\rm rms}$, is plotted. Each halo in Figure \ref{fig:2d_significance} shows a large diversity of planes. This plot is indicative of the fact that there is no uniquely defined plane. For a given number of satellites in the plane there exist a variety of possible thicknesses. And furthermore it is not clear if choosing one satellite less in the plane and thus decreasing its thickness is ``better" than choosing a higher number of satellites in the plane. Therefore it is a priori difficult to decide or to define which ones of the planes should be taken as the ``best" plane. We like to mention, that except for one low concentration halo only high concentration haloes host planes which contain more than 13 satellites. This problem was also investigated by \citet{Cautun2015}, who introduced the ``prominence" of a plane, the inverse probability of obtaining by chance a plane of given number of satellites with a thickness thinner than some predefined value. \citet{Cautun2015} used the ``prominence" to decide which plane is the ``best" to choose. However, for this work we decided to choose the ``best" plane for a given halo by a slightly simpler criteria. We choose the plane with the highest number of satellites as the ``best" plane. If there are two of such planes for a given halo we choose the thinner one of these planes as the ``best" plane.

Figure \ref{fig:planes} shows a visual impressions of some of the planes found by this algorithm. Satellites in the plane are shown as colored dots with the color indicating the sign of the line-of-sight velocity and arrows in each plot indicate the other two velocity components. In this way, the kinematical coherence of the planes inferred from the line-of-sight velocity is immediately visible by eye. For example, the upper right panel clearly shows a plane which appears to be rotating to some degree, whereas a look at the edge-on view shows that at least 3 satellites ($\sim$20\%) have a very high velocity component perpendicular to the plane. This means these satellites are likely to be chance alignments, which will quickly leave the plane. Nevertheless, this plane shows similar features compared to the Andromeda plane of satellites with 15 satellites in the plane and 13 co-rotating ones. Looking at the face-on view (right panel), it also shows a high degree of lopsidedness, with nearly all of the satellites on one side of the host center (more satellites on the upper half of the plot). The projected radial rms value parallel to the plane is $\sim127$ kpc, higher than the calculated value for Andromeda (compare sect. 3.1).
The other planes in this figure do not only have less satellites in the plane but also show less coherent spatial and kinematical planes. The two lower right panels do not show rotating planes, and the corresponding edge-on views in the left panels show a higher fraction of satellites with high velocity components perpendicular to the plane. 

\section{Spatial analysis of planes}\label{sec:spatial}

\subsection{Plane thickness vs halo concentration}

The most fundamental properties of the plane are its thickness, $\Delta_{\rm rms}$, its radial extension, $\Delta_{\parallel}$ and the number of its constituents, $N_{\rm in}$. As reported in Paper I, the thickness of the plane correlates with the concentration of the host halo. High concentration haloes tend to have thinner planes. One could imagine that this might be due to a more concentrated satellite sample but as Figure  \ref{fig:rad_dist} shows the radial distribution of satellite samples of high concentration haloes is the same as for the other two samples and comparable to the observed one. The fact that high concentration haloes show similar planes compared to the observed one can also be seen as well in the left panel of Figure \ref{fig:rms_per_par}, which shows a comparison of the root-mean-square thickness of the plane with its parallel root-mean-square value, calculated as the rms of the radial distances to the main halo center. 
The corresponding value for Andromeda calculated by \cite{Ibata2013} is indicated by a blue star. The value estimated by \cite{Pawlowski2014} for the parallel rms of the Andromeda plane without the furthest satellite AndXXVII ($r>$ 400kpc) is shown as a blue triangle to give an impression of the uncertainty of the parallel rms value. The grey shaded area indicates a nominal uncertainty range for the parallel and perpendicular rms value resulting from the two values given for the parallel rms value and the $1\sigma$ measurement uncertainty of $0.6$ kpc for the perpendicular rms value. Values for the Andromeda analogue simulations of \cite{Gillet2015} are shown as green stars.\par
The planes found in our simulations span a range from $\sim$ 12 kpc to $\sim$ 17 kpc in perpendicular rms and a range of $\sim$ 110 kpc to $\sim$ 170 kpc in parallel rms extension, comparable to the observed plane around Andromeda and the planes found by \citet{Gillet2015}. However, the thinnest planes are only found to be associated with the highest concentration (red dots) haloes, and hence the earliest forming haloes, while low and average concentration haloes show slightly thicker planes compared to high concentration haloes and the observations. 

\begin{figure}
\begin{center}
\includegraphics[width=.49\textwidth]{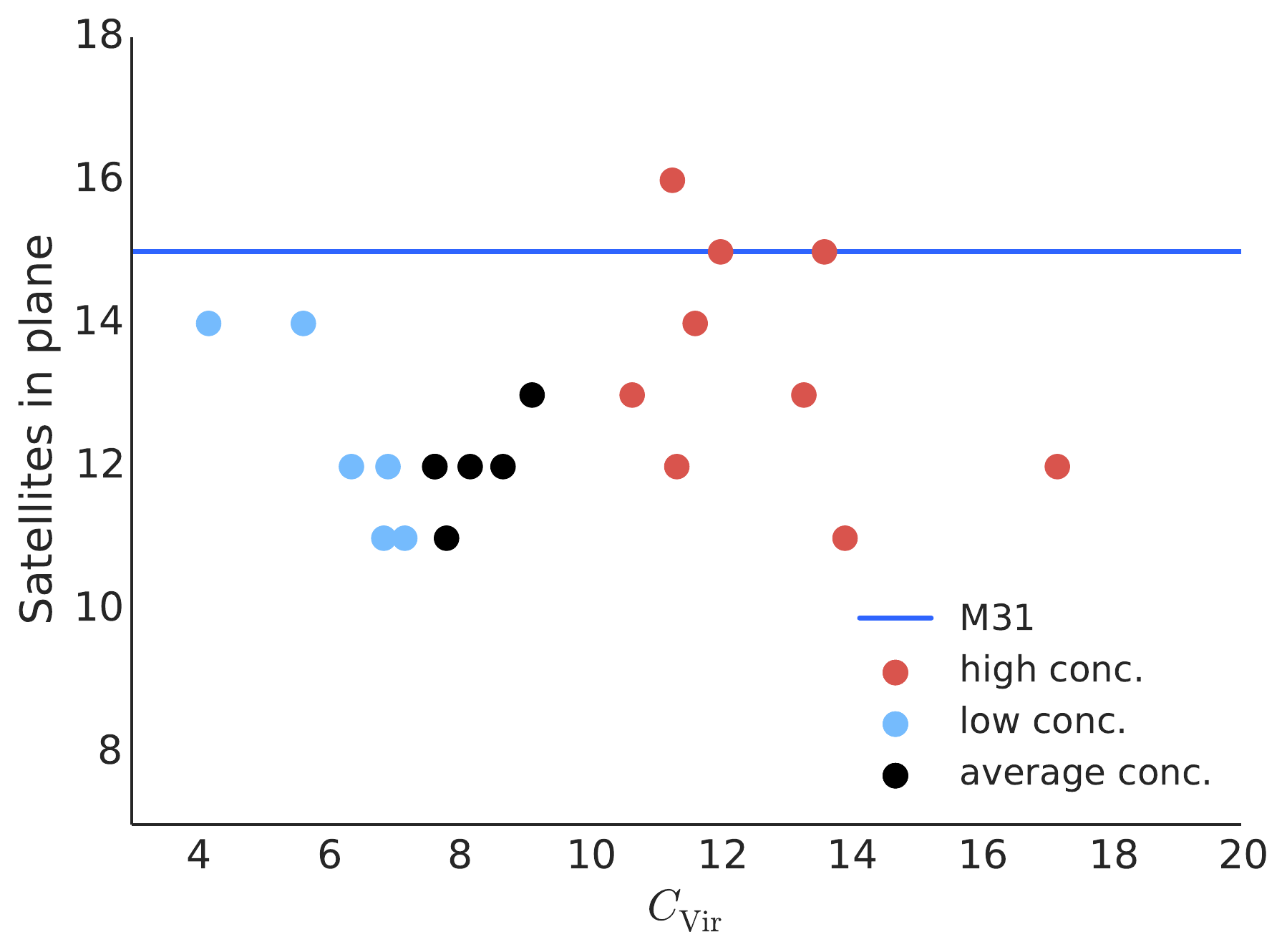}
\end{center}
\vspace{-.4cm}
\caption{The maximum number of satellites in the plane as a function of the concentration. The color-coding divides the haloes into high (red), average (black) and low (blue) concentration haloes. The \emph{blue line} shows the value of 15 satellites in the plane obtained for Andromeda.}
\label{fig:abs_num}
\end{figure}

One can clearly see that the planes in the simulations do not show the same high radial root-mean-square value as observed by \citet{Ibata2013} around Andromeda. But this is no major problem since the root-mean-square value can be biased by outliers with very large radial distances from Andromeda, such as the satellite AndXXVII  with radial distance further away from the center of Andromeda than 400 kpc (larger than the virial radius of Andromeda and the simulated haloes). Excluding AndXXVII, the radial rms drops significantly from 191 kpc to 150 kpc. Therefore, it might be more meaningful to compare the two-dimensional rms value of the radii of satellites in the plane as projected on the sky, as this is the actual observed value. 

The right panel of Figure \ref{fig:rms_per_par} shows the 2-dimensional root-mean-square thickness parallel and perpendicular to the richest planes found when projected on the plane of the sky. The mean values of planes for individual haloes are shown as dots, color-coded by halo concentration. In comparison to the radial root-mean-square value shown in the left panel of Figure \ref{fig:rms_per_par} these values are in better agreement with the observed value and most of the high concentration haloes are consistent with the observed values. Furthermore when comparing the simulations to the observations one has to keep in mind that the observations include satellites obviously outside the virial radius of Andromeda. 
This selection of satellites biases the 3-dimensional root-mean-square value of the observations high compared to the simulations where this selection was not applied for several reasons. The most important reasons is that if this selection criteria was used for the simulations one would have to correct for every viewing angle the number of the 30 most massive satellites since some would be excluded by this criteria. For example, the 3-dimensional root-mean-square extension of the plane including all 15 satellites is $\Delta_{\parallel}=191.9$ kpc, excluding the 2 satellites outside the virial radius of Andromeda (AXVI, AXXVII) the value is calculated to be $\Delta_{\parallel}=129.6$ kpc. In contrast to that, the projected 2-dimensional rms extension of the plane is less effected by this selection ($\Delta_{\parallel 3d}=96$ kpc vs. $\Delta_{\parallel 2d}=97$ kpc)

\begin{figure}
\begin{center}
\includegraphics[width=.49\textwidth]{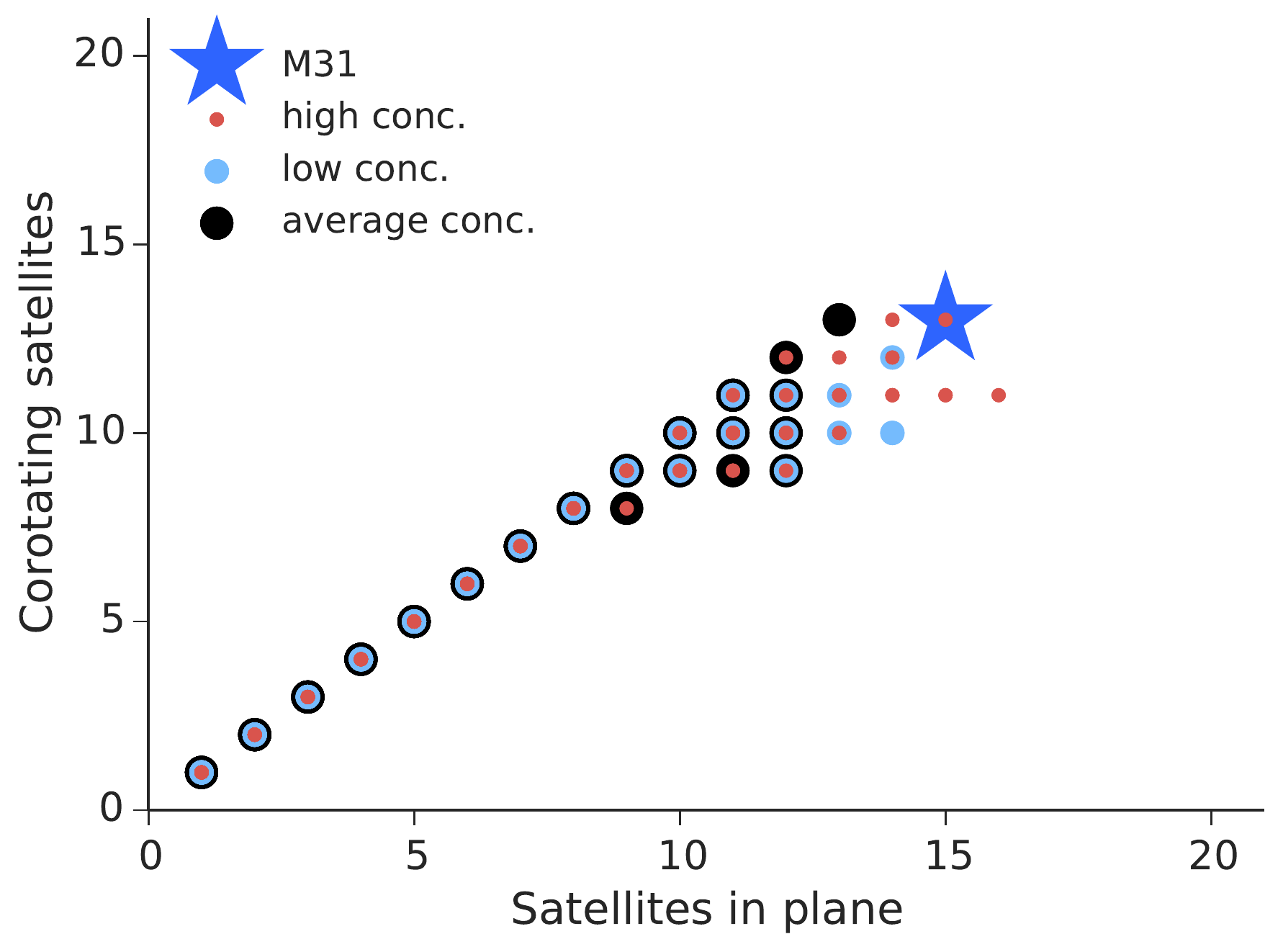}
\end{center}
\vspace{-.4cm}
\caption{The maximum number of co-rotating satellites vs. the number of satellites in the plane, color-coded by halo concentration. The big blue star shows the observed values of Andromeda's plane with 15 satellites in the plane and 13 co-rotating ones. Big black dots show values for planes found in average concentration haloes, middle blue dots show the same for low concentration haloes and small red dots show the results for high concentration haloes.}
\label{fig:abs_num_rot}
\end{figure}

\begin{figure*}
\begin{center}
\includegraphics[width=\textwidth]{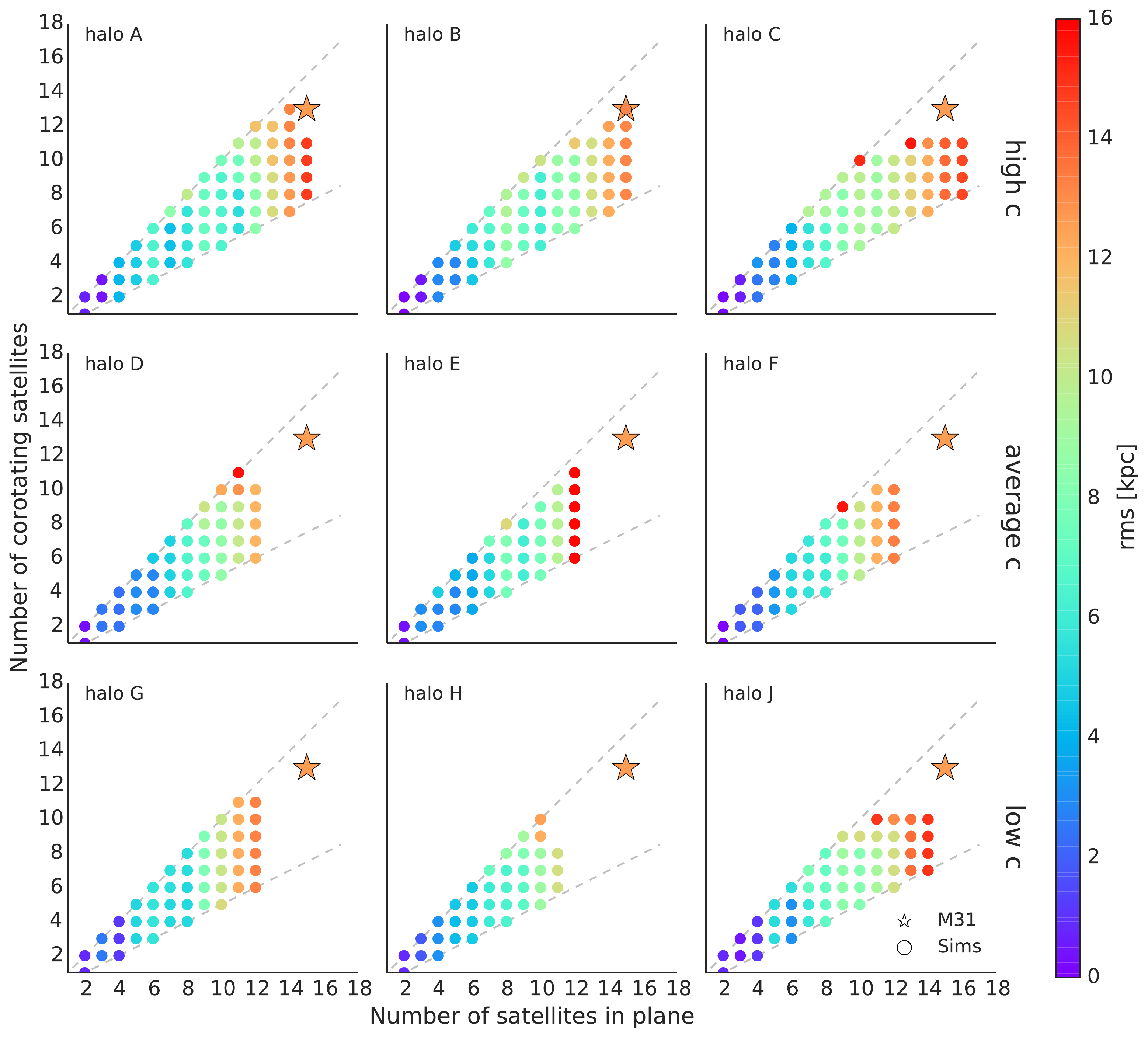}
\end{center}
\vspace*{-.5cm}
\caption{The number of co-rotating satellites vs. the number of satellites in
  the plane for a selection of haloes. The points are color coded by
  the rms thickness ($\Delta_{\rm rms}$)  of each plane. The
  \emph{star} marks the values observed for Andromeda (15 in the plane,
  13 co-rotating). The \emph{dots} show the values of different planes found per simulated halo. The \emph{top row} shows the high concentration haloes, the \emph{middle row} the average concentration haloes and the \emph{bottom row} the low concentration haloes. The dashed grey lines show the physical limits of 100\% and 50\% co-rotation respectively.}
\label{fig:corot_grid}
\end{figure*}

\subsection{Number of satellites in the plane}

As was shown in Paper I and can be seen from Figure \ref{fig:rms_per_par}, for a given number of satellites in a plane, the thickness of the plane correlates with the concentration of the main halo. Furthermore from Figure 2 of Paper I it is obvious that the thickness of the plane is also strongly dependent on the number of satellites in it. Together with the results from Figure \ref{fig:rms_per_par} in the previous section the question arises whether, for a given maximum root-mean-square thickness, the number of satellites in the plane correlates with the concentration as well.
Figure \ref{fig:abs_num} shows the maximum number of satellites in a plane as a function of the concentration. This plot shows, that there is, if at all, only a weak trend with concentration with an overall maximum number of about 12 to 13 satellites in a plane, with just three high concentration haloes reaching a satellite count in the plane as high as that observed for Andromeda. But most importantly by showing that high concentration haloes have the same or even higher numbers of satellites in their planes compared to the other planes, this plot confirms that high concentration haloes have indeed thin planes. The thinner planes in high concentration haloes compared to low or average concentration haloes are not due to a more concentrated satellite distribution (as discussed in section 2.2, but see also Appendix A for more tests) nor to a lower number of satellites in the plane or to a smaller virial radius of the high concentration haloes (see Appendix A). Rather, high concentration (and thus earlier forming) haloes seem to produce genuine thin planes. \par 

\section{Kinematic analysis of planes}
\label{sec:kinematic}

\begin{figure}
\centering
\includegraphics[width=\columnwidth]{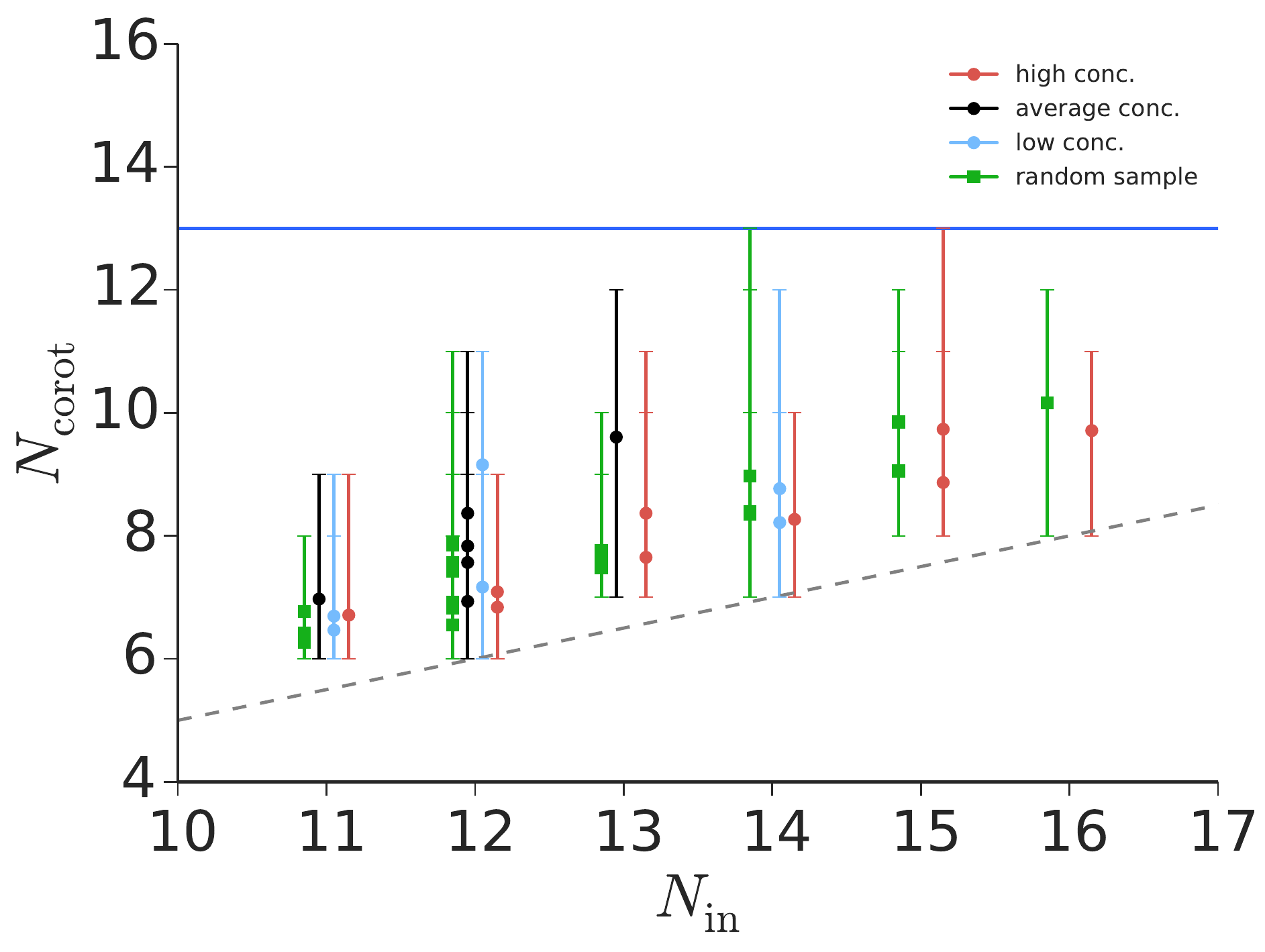}
\vspace*{-.5cm}
\caption{The number of co-rotating satellites counted via the line-of-sight velocity as a function of the number of satellites in the plane. Shown is the mean value for every host in our sample averaged over the full range of viewing angles (180 steps in 1 degree) together with the minimum and maximum value obtained (errorbars). The red dots show high concentration haloes, the black dots average concentration haloes and the blue dots low concentration haloes. The green squares show the result when randomizing the velocities of the satellites in our haloes. The points are slightly offset from each other for better visibility. The \emph{blue line} shows the value of 13 co-rotating satellites measured for Andromeda. The gray dashed line shows the minimum co-rotation fraction of 50\%.}
\label{fig:rot_frac_angle}
\end{figure}

\subsection{Kinematics of the planes from the line-of-sight velocities}
\label{sec:los-kinematics}

In order to investigate the co-rotation of the planes obtained from the plane finding algorithm, for every possible number of satellites in the planes, $N_{\rm in}$, we calculate the number of co-rotating satellites, $N_{\rm corot}$. For this purpose, like for the observations, we use the sign of the line-of-sight velocity in an edge-on view as a proxy for the co-rotation.
Figure \ref{fig:abs_num_rot} presents the results for all main haloes in our sample. Many of the haloes give the same result leading to filling the same points in the diagram.
Up to values of about 10 satellites in the plane, there are viewing angles from which the plane looks like a fully rotating one (with 100\% co-rotating). Interestingly, there is only a slight correlation between the number of co-rotating satellites and the halo concentration. And as was found before, there is no clear dependence of the absolute number of satellites on the halo concentration. There is quite wide scattering among haloes of the same concentration, but again, only high concentration haloes exhibit the highest numbers of satellites in their planes and the highest number of co-rotating satellites.\par  

Despite haloes with very high inferred co-rotation fractions the plane finding algorithm also returns planes which have a much lower inferred co-rotation fraction. To provide a better overview over the possible plane configurations, Figure \ref{fig:corot_grid} shows the outcome of the plane finding algorithm for a selection of 9 haloes of different concentrations. For every value of the number of satellites in the plane $N_{\rm in}$,
the plots show the number of co-rotating satellites, $N_{\rm  corot}$. Every dot in the plot represents a different plane. The points are color-coded according to the thickness of the plane $\Delta_{\rm rms}$. This plot is summarizing the key parameters of the planes ($N_{\rm in}$, $N_{\rm corot}$ and the thickness $\Delta_{\rm rms}$), and it shows again that there is some arbitrariness in selecting the best plane. In particular, it shows that for a given number of satellites in the plane, there is always a plane or viewing angle for which there is no kinematic coherence (50\% co-rotating). As it was concluded from the previous plot, for all of the haloes one can find planes with up to about 10 members that have a 100\% co-rotating fraction. As would be expected, planes consisting of more satellites tend to be thicker. The thickness of the plane, to first order, is also independent of the co-rotation fraction, except for at the highest values of co-rotation fractions.
  
A comparison between haloes of different concentrations reveals that only high concentration haloes show planes comparable to the observed ones (see also Figure \ref{fig:abs_num_rot}), with 15 satellites in the plane and a comparable high co-rotation fraction while at the same time showing a similar low thickness. Haloes from the average or low concentration sample fall short in number of satellites in the plane with only about 12 satellites in the plane. Furthermore these haloes show at fixed number of satellites in the plane thicker planes compared to the high concentration sample.  

\begin{figure*}
\centering
\includegraphics[width=.49\textwidth]{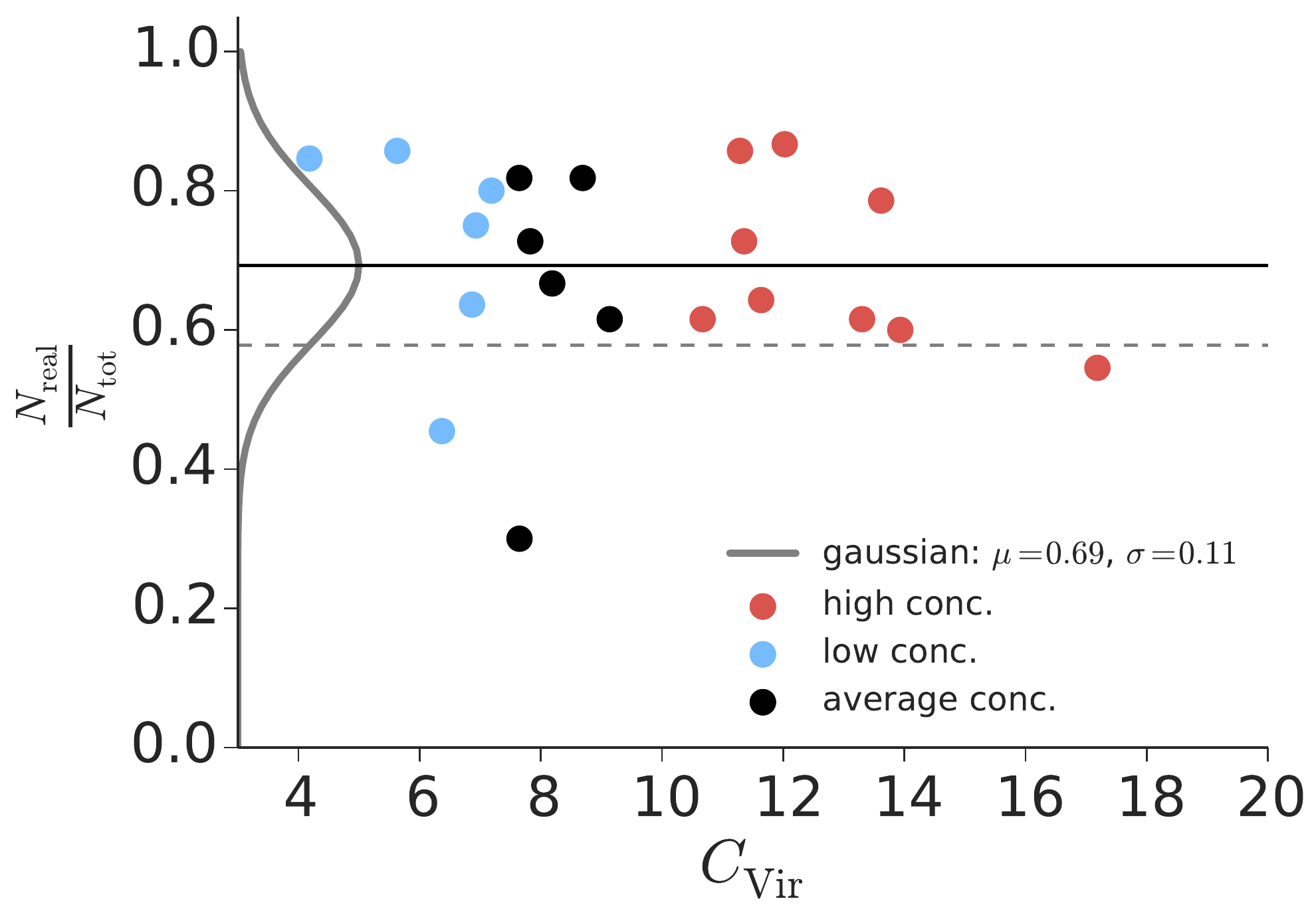}
\includegraphics[width=.49\textwidth]{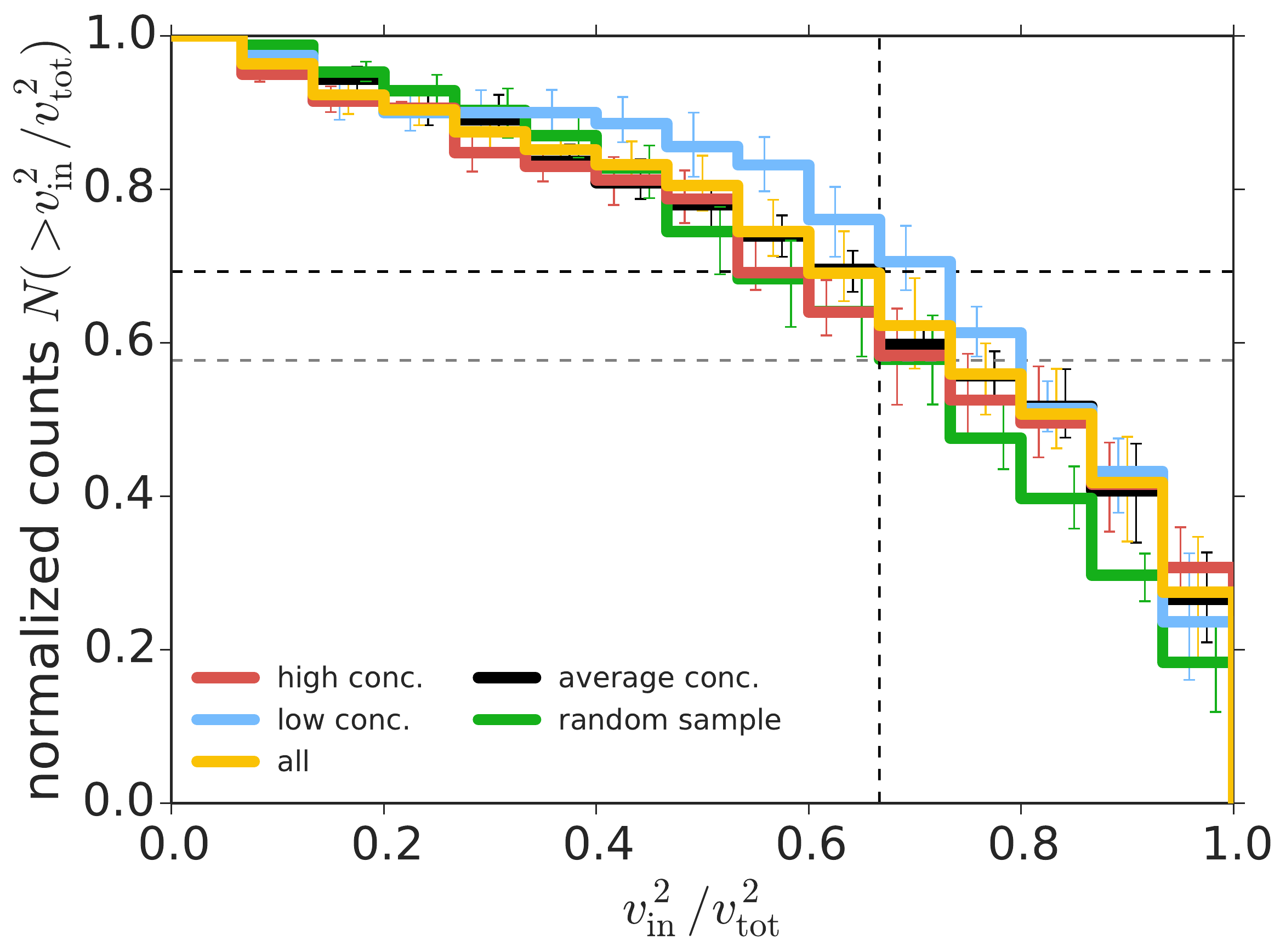}
\vspace*{-.25cm}
\caption{\emph{Left panel}: The fraction of satellites kinematically bound to the plane, color-coded by concentration. The selection criteria for the satellites to be kinematically confined in the plane is the ratio of the velocity component in the plane to the total velocity: $v_{\rm plane}^2/v_{\rm tot}^2>\frac{2}{3}$. The solid black line indicates the mean value of $(69\pm11)$\% bound satellites and the dashed gray line shows the expectation of 57.7\% from a sample of satellites with randomized velocities. \emph{Right panel}: Cumulative probability distribution of $v^2_{\rm plane}/v^2_{\rm tot}$ for the high, average and low concentration sample (red, black, blue lines) as well as the whole sample of haloes (yellow line) and the same sample with randomized velocities (green line). The black dashed lines show the expectation for $v^2_{\rm plane}/v^2_{\rm tot}>2/3$ of $(69\pm11)$\% and the gray dashed line shows the value of $57.7$\% expected for a random sample. The errorbars are slightly offset from each other for better visibility.}
\label{fig:real_planes}
\end{figure*}

\begin{figure}
\includegraphics[width=\columnwidth]{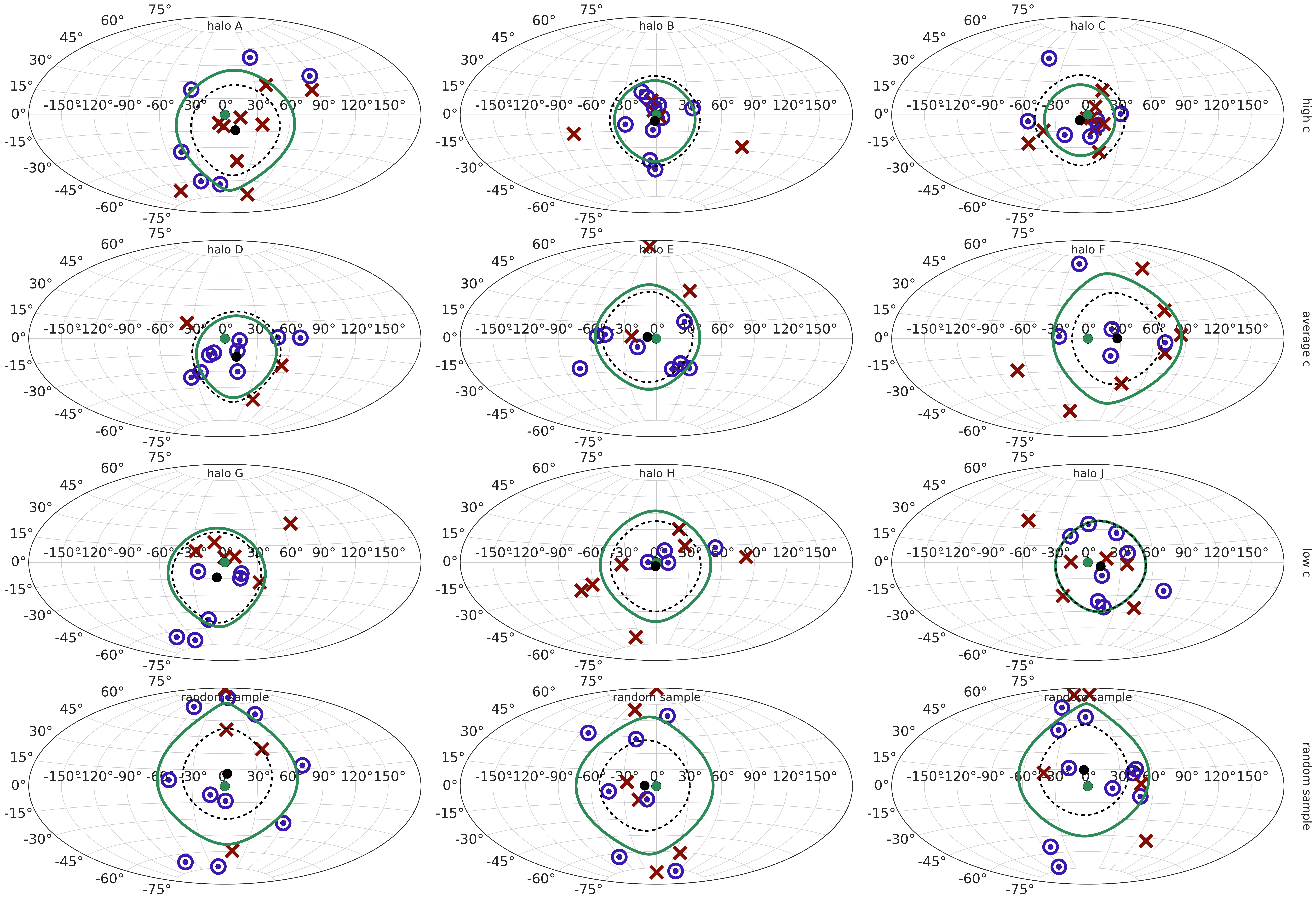}
\includegraphics[width=\columnwidth]{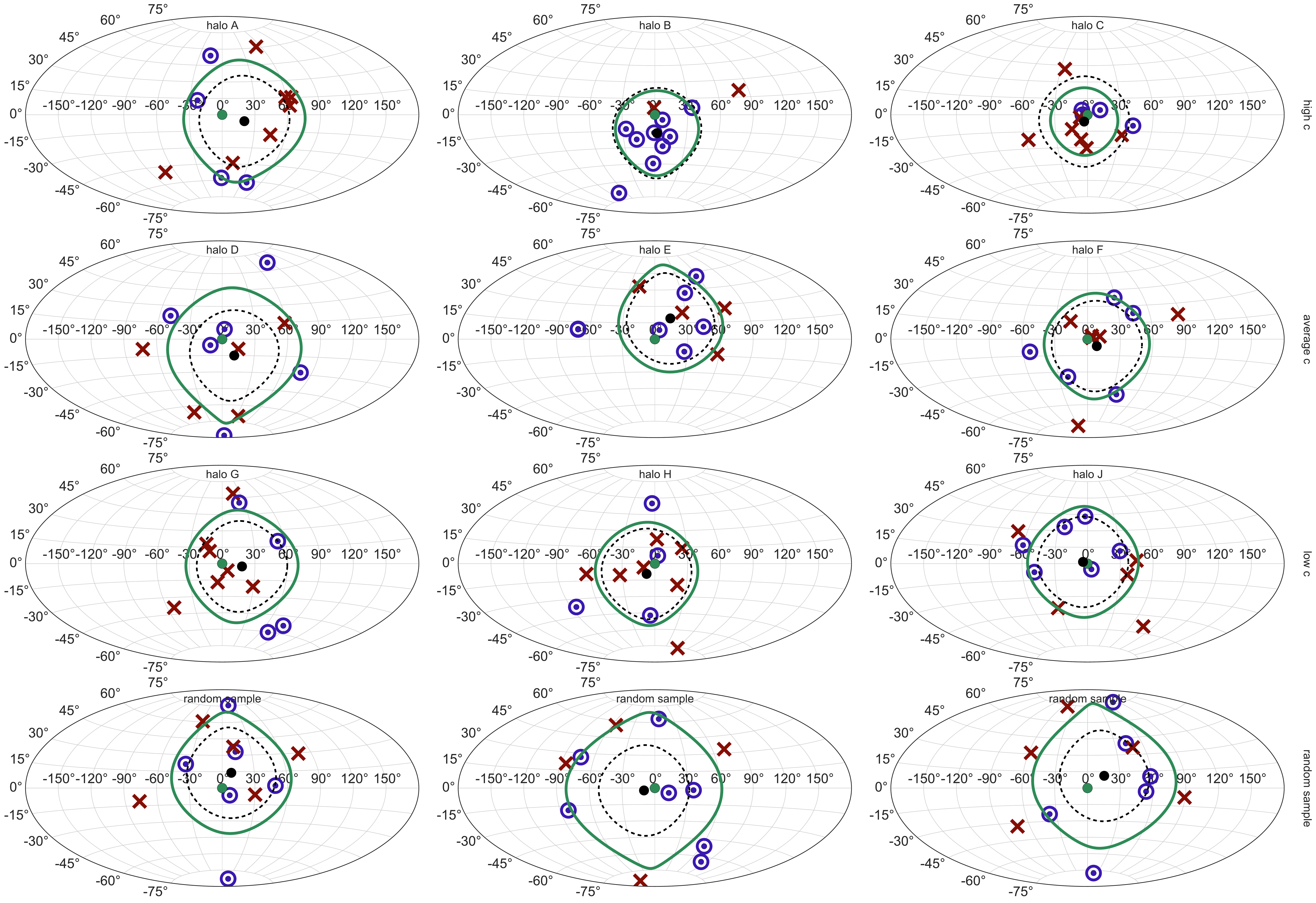}
\includegraphics[width=\columnwidth]{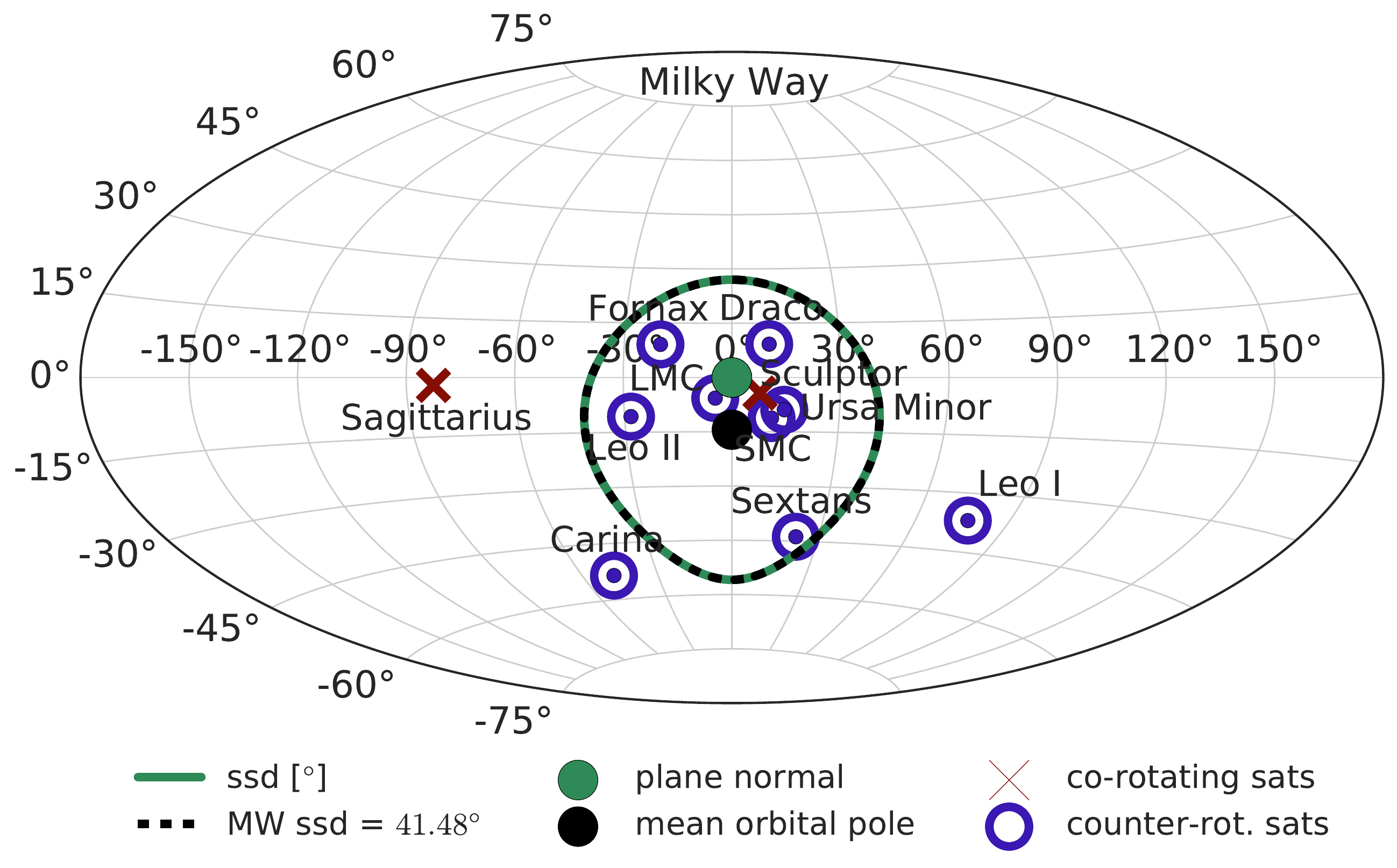}
\vspace*{-.6cm}
\caption{A Hammer-Aitoff projection of the orbital poles of the satellites. \emph{The Upper panel} shows the satellites in the plane of halo B, the \emph{middle panel} shows the 11 most massive satellites of halo B and the \emph{lower panel} shows the 11 classical Milky Way satellites with known proper motions (Table 2, \protect\cite{Pawlowski2013}). The \emph{green dot} shows the direction of the plane normal and the color-coded symbols show the orbital poles of the co-rotating (red crosses) and counter-rotating (blue circles) satellites, respectively. Since angular momentum vectors are axial quantities, we invert the angular momentum vectors of clockwise orbiting satellites to scatter around $\phi=0^\circ$ rather than $\phi=\pm180^\circ$ and show them as blue circles with a dot in order to provide a better comparison of the clustering of orbital poles around the plane normal. The orbital poles for counter-clockwise rotating satellites are kept fixed and are shown as red crosses. The \emph{green solid line} shows the spherical standard distance calculated from \emph{all} the satellites in the plane with the \emph{black dot} showing the average orbital pole of all of the satellites. For all the Milky Way satellites we obtain a value of ssd=$41.5^\circ$ which is shown with a \emph{black dashed line}.}
\label{fig:ang_mom_obs}
\end{figure}

\subsection{Number of co-rotating satellites for different viewing angles}

Once a plane is identified there are many possible line-of-sights view the plane edge-on. Thus we are interested in inferring the co-rotation for the same plane from different viewing angles by rotating it around its normal vector and measure the number of co-rotating satellites via the sign of the line-of-sight velocity. Since the line-of-sight velocity represents only one component of the three-dimensional velocity of the satellites on their orbit around the halo centre we expect a dependence on particular viewing angles. If the planes were fully co-rotating with satellites on circular orbits, there should be no dependence of the line-of-sight co-rotation fraction on the viewing direction. Thus, when rotating the plane around the normal vector, the line-of-sight count of co-rotating satellites should not vary. But the orbits of the satellites are not perfect circles and additionally there could be interlopers in the plane that only align with it by chance. Therefore the line-of-sight velocity count should depend on the exact viewing direction. Figure \ref{fig:rot_frac_angle} shows the number of co-rotating satellites counted via the line-of-sight velocity as a function of the number of satellites in the plane. We show in this plot the mean value of co-rotating satellites together with the minimum and maximum value obtained. The first thing to notice is that there is quite a large variation in the number of co-rotating satellites as the viewing angle changes. The mean number of co-rotating satellites is around 7 to 10 (60\% co-rotation fraction) with a scatter of about 2 to 3. Really high co-rotation fractions are only reached at some peaks of special viewing angles and in only about 10\% of all viewing angles a high value of co-rotating satellites is reached.

We recover that high concentration haloes have higher numbers of satellites in the plane and thus reach higher co-rotation fractions. But even these haloes do not exhibit fully rotating planes. The reason for the large variance in co-rotation fractions is that the planes seem to be not fully kinematically coherent structures but include a significant amount of interlopers, which have just joined the plane coincidentally. 
This can be concluded from doing the same analysis but randomizing the velocities of satellites in the plane of our hosts. The line-of-sight counting of co-rotating satellites for a random sample exhibits the same features (mean number, scatter) as obtained for the simulated planes. Therefore we conclude that inferring the kinematical coherence of the plane from only a measurement of the line-of-sight velocity is not robust, even a sample of satellites with random velocities can appear as apparently rotating as Figure \ref{fig:rot_frac_angle} shows.

The conclusion of a significant fraction of interlopers in the plane is further supported if one compares the 2-dimensional velocity component in the plane $v_{\rm in\ plane}$ to the total velocity $v_{\rm{tot}}$ of satellites. Assuming, the plane normal points in the $z$-direction, one obtains:
\begin{equation}
v_{\rm plane}=\sqrt{v^2_{x}+v^2_{y}}\quad\quad
v_{\rm tot}=\sqrt{v^2_{x}+v^2_{y}+v^2_{z}}
\end{equation}
If the motion of satellites would be fully random and isotropic, one would expect every velocity component to have the same magnitude and the above comparison would result in $v^2_{\rm plane}/v^2_{\rm tot}=2/3$. However this is obviously not the case for a \lcdm cosmology where satellite orbits show some kind of ordered motion. 
Nevertheless, the isotropic case can be used to discriminate between ordered motion in a plane and unordered motion. Therefore we use this analysis to estimate the fraction of satellite rotating in the plane. If $v^2_{\rm plane}/v^2_{\rm tot}<2/3$, the velocity component perpendicular to the plane is higher than in a random and isotropic case and the satellite can be regarded as not taking part in the motion in the plane. Vice versa, if $v^2_{\rm plane}/v^2_{\rm tot}>2/3$, the 2-dimensional velocity component in the plane is higher than in the random and isotropic case, and the satellite`s motion can be regarded as motion in the plane. For a sample of satellites with random velocities one would then expect that 57.7\% would fulfill the above criteria. This can be estimated from the area of a spherical cap of height $(1/3)^{0.5}$ divided by the area of a hemisphere if one assumes the velocity components to be points on the surface of a sphere.

The left panel of Figure \ref{fig:real_planes} shows the fraction of satellites belonging to the plane if the above criteria is used to discriminate between kinematic members and interlopers as a function of concentration. The simulations show a fraction of $\sim$30\% interlopers or vice versa a fraction $\sim$70\% of the satellites beeing kinematically bound to the plane according to the above criteria with no much difference between the high and low concentration haloes. This fraction is significantly higher then expected for a random sample and thus implies that \lcdm produces some degree of motion in a plane although not all the satellites are necessarily confined to the plane. These results are in agreement with the findings of \cite{Gillet2015}, who find a fraction of about 1/3 of the satellites to be interlopers. This conclusion is also supported by the analysis of the angular momentum vectors of the plane members presented in the next subsection as well as the analysis of the past orbits of the satellites making up the plane at the present time.

The right panel of Figure \ref{fig:real_planes} shows the cumulative probability distribution function for $v^2_{\rm plane}/v^2_{\rm tot}$ for both \lcdm haloes (yellow line) and a random sample (green line). The random sample is created from the \lcdm sample by randomizing the velocities and keeping the other parameters of the satellites fixed. Comparison of both samples shows that there is a trend for \lcdm to prefer rotation in a plane with slightly higher probabilities for values of $v^2_{\rm plane}/v^2_{\rm tot}>0.5$ compared to the values for the random sample. Although for intermediate values of $v^2_{\rm plane}/v^2_{\rm tot}$ there is a difference apparent between the individual concentration samples, they are within their error bars consistent. The difference might be mainly due to a smaller number of satellites in the plane for low concentration haloes.

\begin{figure*}
\center
\includegraphics[width=\textwidth]{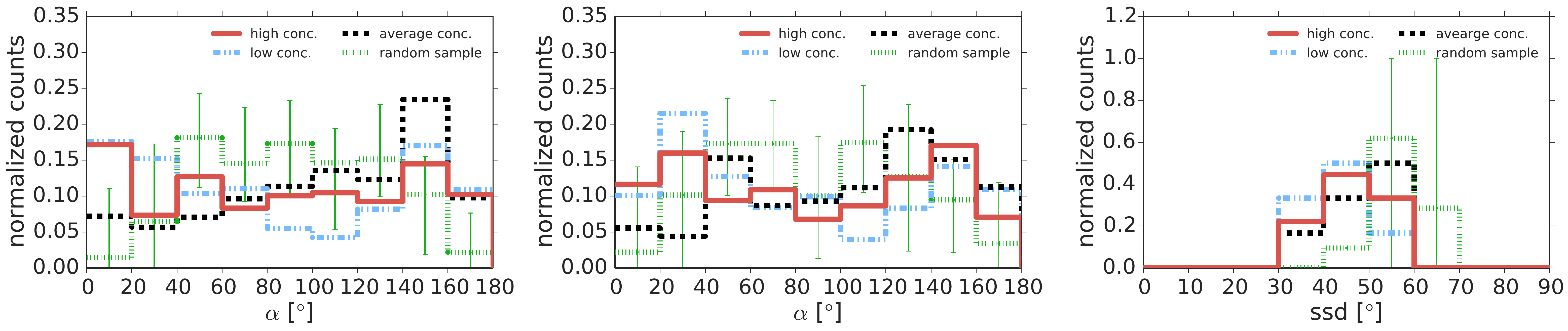}
\vspace*{-.5cm}
\caption{Distribution of the angles between the orbital poles of individual satellites and the plane normal (left panel) as well as the mean orbital pole (right panel) for the three different concentration samples. High concentration haloes are shown with a red solid line, average ones with a black dashed line and low concentration haloes with a blue dotted-dashed line. Additionally we show the same for our sample of hosts with randomized velocity vectors of the satellites as a green dotted line. The scatter in the different samples is almost the same, therefore we only show the scatter for the random sample as a reference.}
\label{fig:hist}
\end{figure*}

\subsubsection{Angular momentum analysis}
\label{sec:ang_mom}

In the previous analysis of the planes we closely followed the methods used to quantify the co-rotation of the plane of satellites around Andromeda. Thus, only a one-dimensional measure of the kinematical coherence of the plane, the line-of-sight velocity, has been used. In order to capture the motion of the satellites in the full 3-dimensional case, and not only as a projection by counting the number of co-rotating satellites by their line-of-sight velocity, we measure the angular momentum vector of each satellite in the plane and compare its direction to the direction of the plane normal and the mean angular momentum vector of all the satellites in the plane (mean orbital pole). For this analysis, the plane normal vector is set to $(\phi,\theta)=(0,0)$ in spherical coordinates as seen from the halo center. The angular momentum vector of the satellites that are fully orbiting counter-clockwise in the plane should point in the same direction as the plane normal, while for clockwise orbiting satellites the angular momentum vector is pointing in the opposite direction, hence to $(\phi,\theta)=(-180,0)$ or $(\phi,\theta)=(180,0)$. Therefore, the main difference between the co- and counter-rotating satellites is the direction of their angular momentum vectors. Satellites that are not orbiting within the plane show a deviation in the directions of their orbital poles from the plane normal.

A similar study was done by \cite{Pawlowski2013} for the 11 classical Milky Way satellites. This study revealed that the angular momentum vectors of 8 of these satellites are clustered. One measure of the clustering of orbital poles is given by the spherical standard distance of the orbital poles. The spherical standard distance is the scatter of the orbital poles in spherical coordinates (see \cite{Pawlowski2012} for a definition of their ssd). \cite{Pawlowski2013} obtained a value of ssd=$29.3^\circ$ for the 8 most clustered Milky Way satellites. We decide to include all the 11 satellites into our analysis and use a slightly different definition of the ssd given by the following formula:
\begin{equation}
\rm{ssd}=\sqrt{\frac{\sum_{i=1}^{k}\left[\rm{arccos}\left(|\left<\boldsymbol{n}\right>\cdot\boldsymbol{n}_{i}|\right)\right]^2}{k}},
\end{equation}
where $\boldsymbol{n}_i$ are the angular momenta vectors of the satellites and $\boldsymbol{\left<n\right>}$ is the mean angular momentum vector of the satellites in the plane (calculated before inverting them). The difference to \cite{Pawlowski2013} is that we take the absolute value of $|\left<\boldsymbol{n}\right>\cdot\boldsymbol{n}_{i}|$ and we choose to include all 11 satellites instead of only 8. We reproduce their Figure 1 from the values given in Table 2 of their paper and present it in the lower panel of Figure \ref{fig:ang_mom_obs} with our color coding convention. Calculation of the spherical standard distance for all 11 Milky Way satellites with our formula gives a value of ssd=$41.5^\circ$, somewhat higher than for the 8 most clustered ones with ssd=$29.3^\circ$. The same Figure also shows a visual comparison between the distribution of orbital poles of the satellites in the plane using our plane finding algorithm (upper panel) and using our plane finding algorithm only on a sample of the 11 most massive sub-haloes at infall time (middle panel). There is not much difference between the two different selection methods when comparing the distribution of orbital poles. The ssd is almost the same for both cases and compares well to the Milky Way plane. Therefore we focus only on the planes discussed before.

\begin{figure}
\center
\includegraphics[width=\columnwidth]{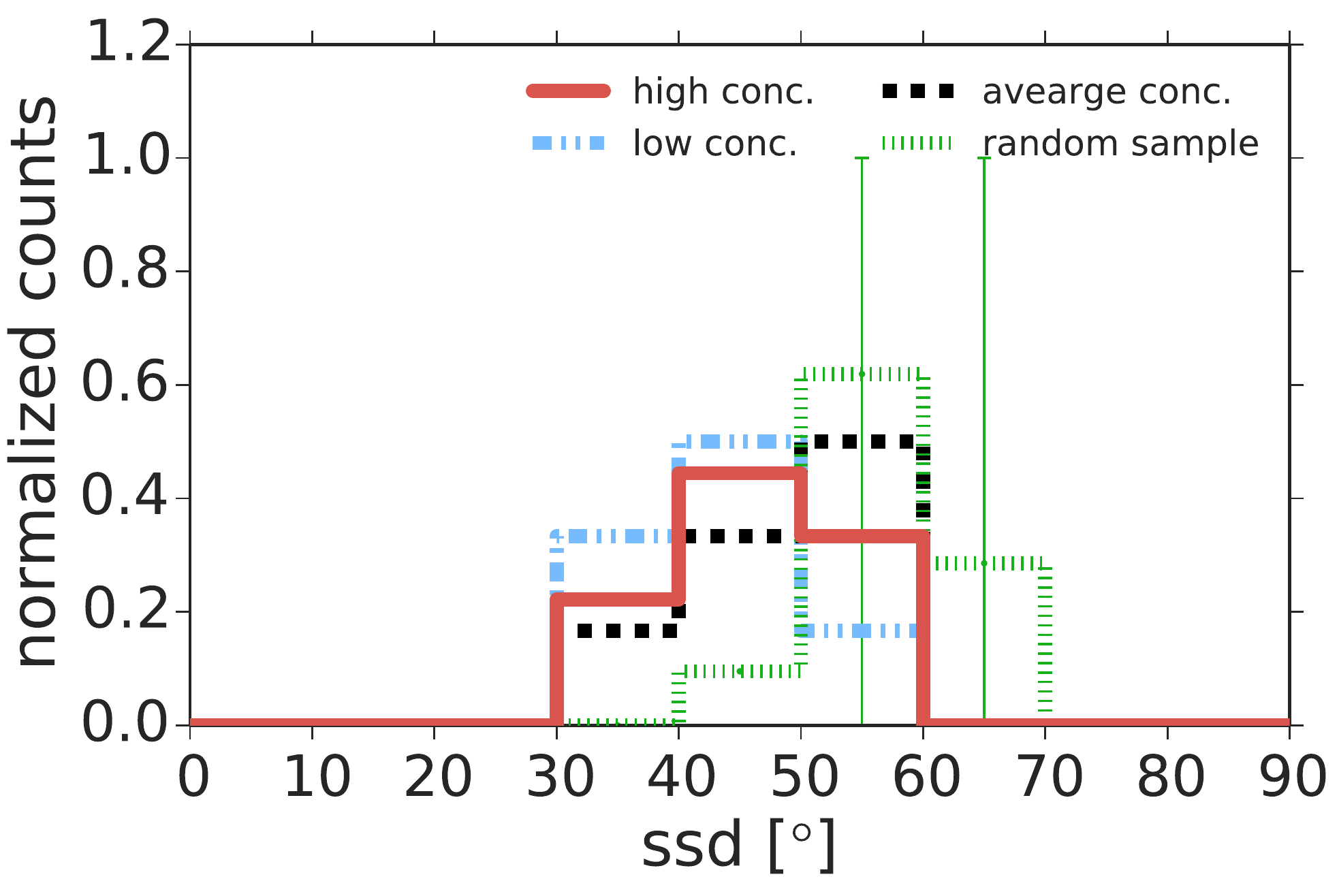}
\vspace*{-.5cm}
\caption{Distribution of values for the spherical standard distance for different concentration samples with the same color coding and linestyle convention used before. The green bars show the distribution of ssd for the whole sample of hosts with randomized velocity vectors. Again we show only the scatter for the random sample as reference.}
\label{fig:ssd}
\end{figure}

There is a large degree of scattering in the angular momentum directions around the direction of the plane normal, indicating non-ordered motion. If the planes were indeed to some degree rotating structures as the line-of-sight velocity measurement would indicate, more or less all of the angular momentum directions would cluster near the plane normal direction at $(\phi,\theta)=(0,0)$. This is indeed true for some of the satellites but a significant amount of satellites shows a large offset from the plane normal/mean orbital pole. To quantify this behavior we show in Figure \ref{fig:hist} the distribution of the angle between the orbital poles of the satellites and the plane normal (left panel) and the mean orbital pole (middle panel) for all 21 hosts. In Figure \ref{fig:ssd} we show the distribution of spherical standard distances for all of our host haloes in comparison to a sample with randomized velocity vectors.

The left panel of Figure \ref{fig:hist} shows that there is not much a difference between the planes found in the haloes of different samples but a strong difference between the \lcdm haloes and the random sample. All \lcdm samples show a trend for the satellites to preferentially align with the plane normal. This is evident from a slight over-abundances of angles with values close to $\alpha\sim0^\circ$ (co-rotating) and close to $\alpha\sim180^\circ$ (counter-rotating). The random sample shows a more uniform distribution of angles with preferred values of $\alpha$ around $90^\circ$. The nearly equal number of satellites with values close to 0 and $180^\circ$ shows that the planes indeed show a spatial alignment but on average do not show a strong kinematical coherence. The same behavior is found for the mean orbital pole which is calculated from the orbital poles of the satellites (right panel of Figure \ref{fig:hist}). Interestingly there is not much difference in the kinematics between the different concentration samples. All hosts show a fraction of about 30\% of the satellites having values of $\alpha\sim90^\circ$ indicating that these satellites do not take part in a motion in the plane. This analysis shows that although \lcdm haloes show a higher degree of co-rotation compared to a random sample these haloes do not exhibit fully kinematically coherent planes. This is further supported by analysing the distribution of spherical standard distance values for the different host haloes. All hosts in our sample show ssd values higher than $30^\circ$ indicative of a large scatter in orbital poles. Although for most haloes, the calculated spherical standard distance is comparable to that of the Milky Way of $\sim41^\circ$, the widespread distribution of orbital poles indicates a large fraction of interlopers. For the haloes with the most widespread distribution of angular momentum directions we expect that there is nearly no kinematical coherence in these haloes, mostly all satellites in the plane will have different orbital planes.
However, comparison to the random sample shows that the scatter is smaller than expected for a sample with randomized velocity vectors. This is in contrast to the result of Figure \ref{fig:rot_frac_angle} where no difference to a random sample was found. Therefore the analysis of orbital poles is better suited to measure the co-rotation of satellites in planes than simply count their line-of-sight velocity. The particular advantage of this measurement is that in contrast to the inference of co-rotation from line-of-sight velocity measurements this method is unique and is not dependend on specific viewing angles. Unfortunately it is hard to obtain proper motion information for most of the satellites in Andromeda to do this analysis (see also next section). It gets even worse if one would do this measurement for satellite planes further away from us than Andromeda.\par

\subsection{Plane evolution and plane stability}
\label{sec:orbit}

With the finding, that the planes of satellites we identified seem to contain a significant amount of satellites not taking part in coherent rotational motion it is interesting to estimate their proper motion. Given the average velocity component of the satellites perpendicular to the plane of $\sim$100 km/s we can estimate their proper motion if they would orbit around Andromeda. Given the distance of the sun to Andromeda of $\sim$780 kpc \citep{McConnachie2012} we estimate the proper motion of the satellites to be $\sim0.03$ mas/yr. This is at the edge of what is currently possible with the HST Promo project \citep{Sohn2012}.
A more elaborate estimate of the plane's lifetime can be made by following the orbit of the satellites in the plane and tracing their positions back in time. This is done in the next subsection for two high concentration haloes that show evidence of kinematically coherent planes (halo A and B). 

\begin{figure*}
\includegraphics[width=\textwidth]{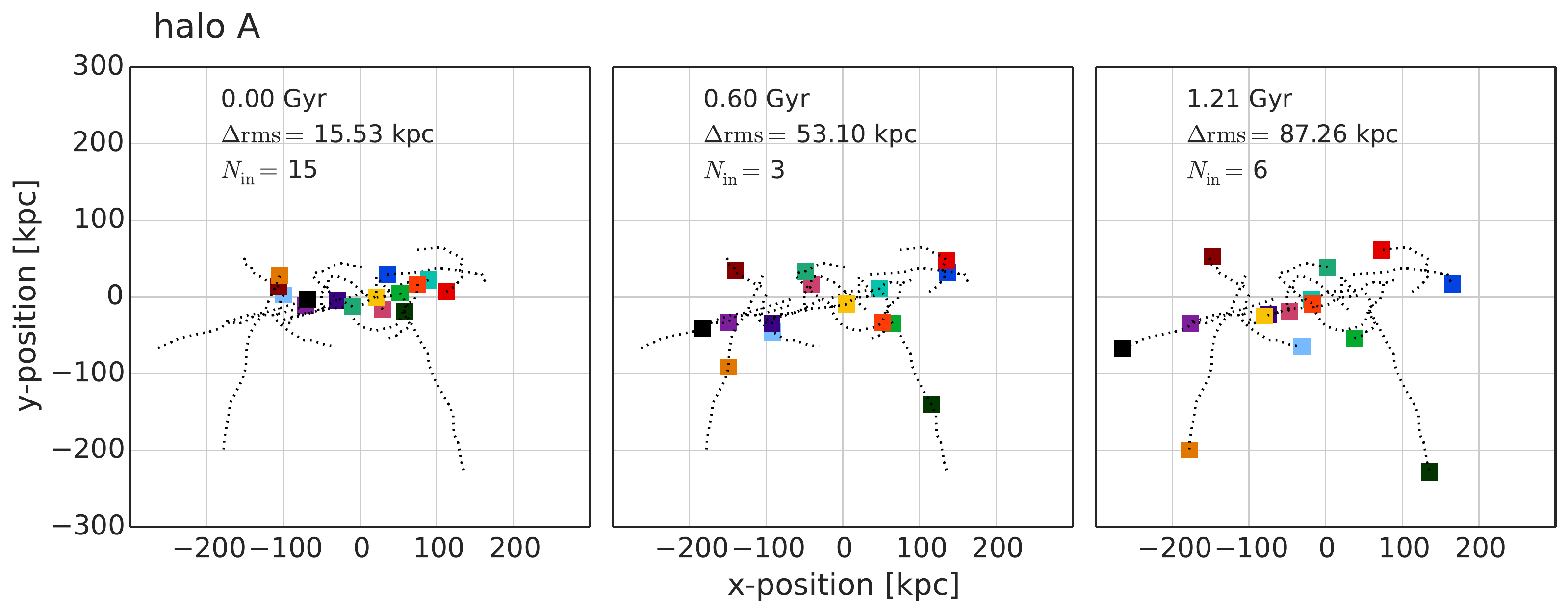}
\includegraphics[width=\textwidth]{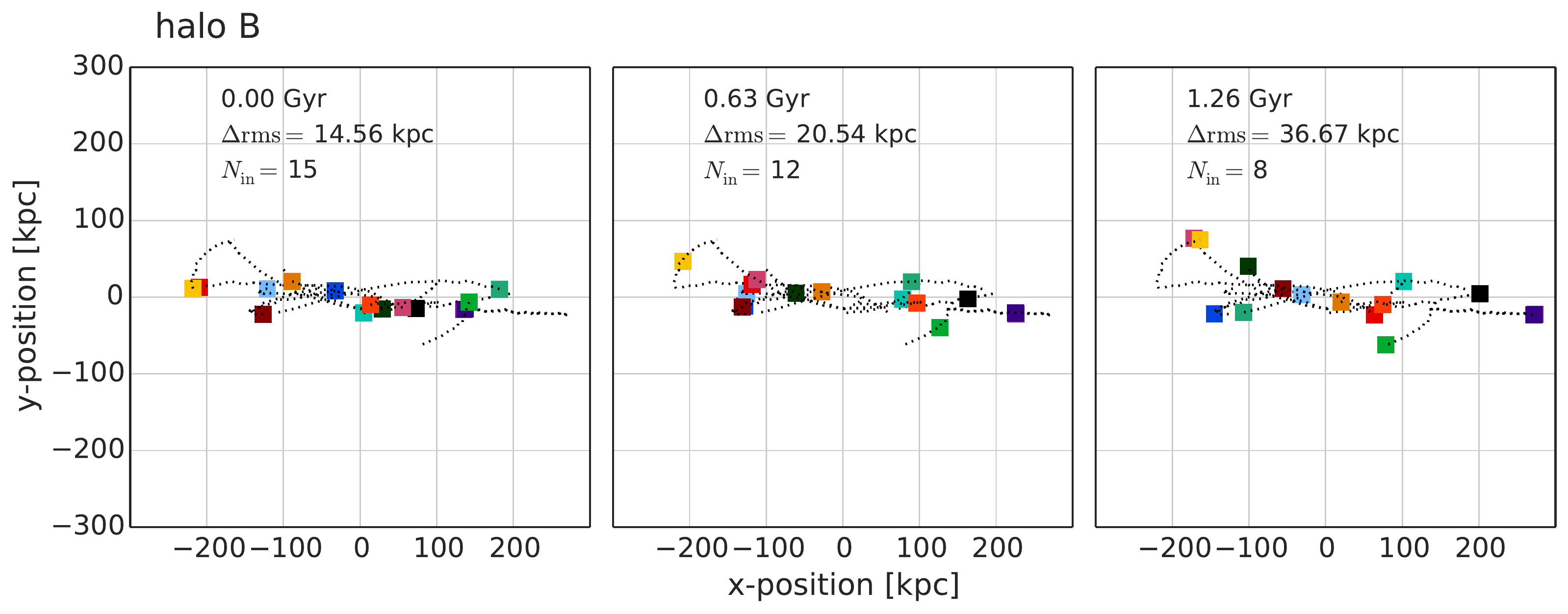}
\vspace*{-.5cm}
\caption{Visual impression of the orbits of the satellites in the planes of halo A (top panels) and halo B (bottom panels) with an edge-on view. \emph{Dashed lines} show the trajectories of the satellites and \emph{colored squares} show the positions of the satellites in the planes. From left to right, snapshots at increasingly earlier times are shown. \emph{Left}: The present day configuration, \emph{middle}: the configuration $\sim$ 0.6 Gyr in the past ($\sim2t_{\rm dyn}$) and \emph{right}: $\sim$ 1.2 Gyr ago ($\sim4t_{\rm dyn}$). In each of the panels, the time, the rms value of the plane and the number of satellites still within the region of the plane at the present time is indicated.}
\label{fig:orbit}
\end{figure*}

\subsubsection{Satellite orbits}

\begin{figure*}
\centering
\includegraphics[width=\textwidth]{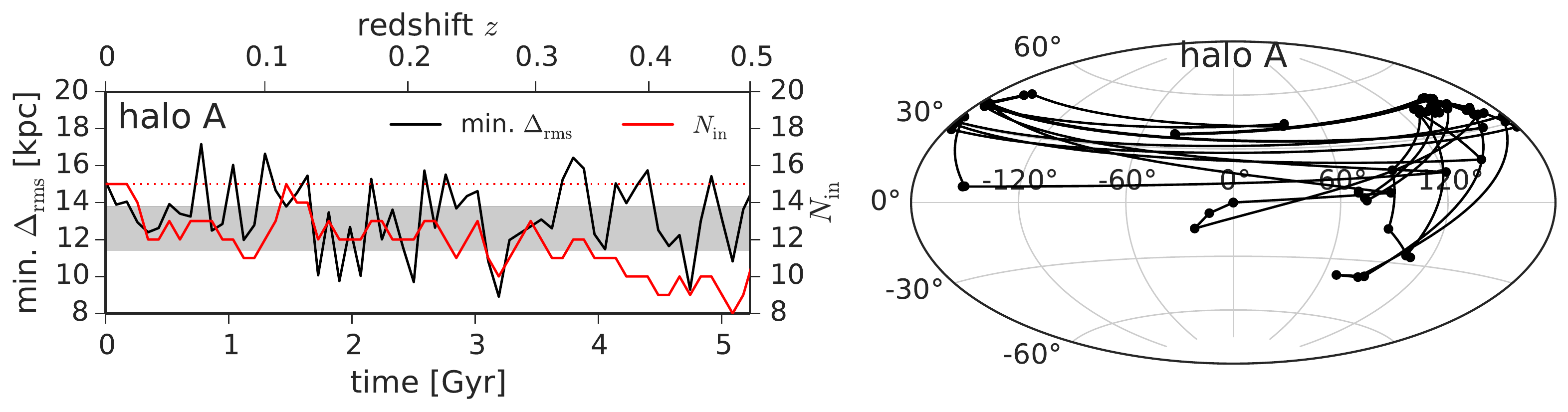}
\includegraphics[width=\textwidth]{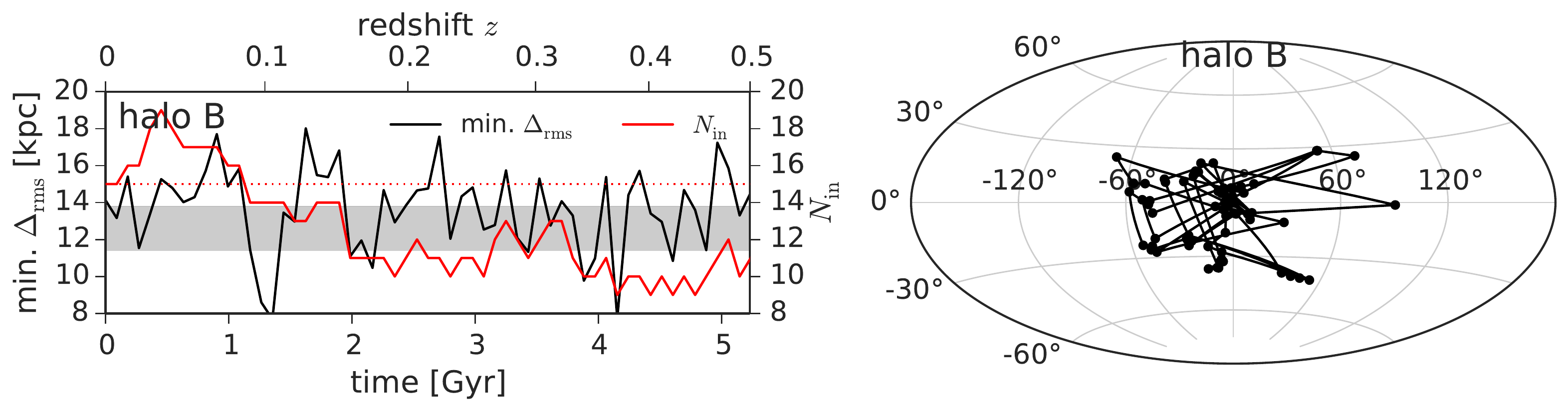}
\vspace*{-.5cm}
\caption{The planes at different time steps in the past. \emph{Left panel}: The rms value perpendicular to the plane (black solid line) and the number of satellites in the plane (red solid line, right y-axis) for different time steps of the simulation down to 5 Gyr in the past. The grey shaded area shows the measurement uncertainty of $\Delta_{\rm rms}= 12.6 \pm 0.6$ kpc of the observed thickness of the plane around Andromeda. The red dotted line shows the number of 15 observed satellites in the plane. \emph{Right panel}: The evolution of the corresponding plane normal in spherical coordinates over the past 5 Gyr. The orientation of the plane was chosen such that at present time the plane normal points in the direction of $(\phi,\theta)=(0,0)$. The \emph{upper panels} show halo A and the \emph{lower panels} show halo B.}
\label{fig:thickening_time}
\end{figure*}

The plane stability and the evolution of its thickness over the last 5 Gyrs can be investigated by tracking the selected 30 satellites back in time and investigating their positions in the past. In order to investigate the plane coherence over time we rerun the plane finding algorithm on the selected satellite samples of 30 satellites at an earlier time to find a new best fitted plane (see Fig. \ref{fig:thickening_time}) probably consisting of a different subset of satellites compared to the $z=0$ best plane to account for satellites presumably moving out of the plane and new satellites joining the plane.

A visual impression of the orbits of satellites (grey dotted lines) in the planes of halo A and B for the last 1.2 Gyr in the past is given in Figure \ref{fig:orbit}. The three panels of the figure also show the position of each of the satellites with colored squares at the present time (left panel), at $\sim$0.6 Gyr (middle panel) and at $\sim$1.2 Gyr (right panel) in the past. From a quick look at the orbits of halo A (upper panels), it is immediately clear that there is almost no motion in the plane, while halo B shows a more coherent motion of the satellites in the plane. For halo A the tracks of the satellites extend up to a height of about 75 kpc above the plane or are in some cases even perpendicular to the plane at the present time. Furthermore, after only 0.6 Gyr the positions of the satellites indicate a plane thickened up to about 50 kpc. For halo B (lower panels) most of the satellite orbits stay close to the plane, while for at least 3 satellites the orbits seem to be nearly perpendicular to the plane at present time. However, the plane of halo B does not show a thickening as pronounced as halo A. After 0.6 Gyr this plane only shows a thickening of a factor of 1.4, up to 20 kpc. After $\sim$1.2 Gyr the plane of halo A is hardly visible, with an rms value of 87 kpc. At this time the plane of halo B is $\sim$37 kpc thick and thus more than doubled its thickness and would by any criterion used for the observations not been regarded as a plane anymore.

This result is confirmed by Figure \ref{fig:thickening_time}, where we show the plane thickness, the number of satellites in the plane and the orientation of the plane normal of halo A and B for many different time steps in the past. Here we restricted the rms value to be smaller than 22 kpc. This figure shows that while the rms value is strongly fluctuating around a value of $\sim$ 14 kpc the number of satellites in the plane is continuously going down such that at redshift $z=0.5$ the plane consists of only 10 satellites. At the same time the right hand side of the plot shows that the orientation of the plane normal is far from constant and does not show a slow precession, which would be expected for a coherent plane. Rather the plane normal jumps around with only slight clustering behavior. Halo A seems to show no temporal coherence of the plane normal, but a slight clustering point at $(\phi,\theta)\sim(130,30)$. One has to keep in mind that the angular momentum vector is a axial quantity and thus the points on the left side of the plot at $(\phi,\theta)\sim(-180,30)$ are basically the same plane as the clustering on the right side of the plot. For halo B the plane normal is somewhat more coherent with a scatter of $\sim30^{\circ}$ and clusters around $(\phi,\theta)\sim(0,0)$ but still shows a significant scatter. However, both of these clustering points are not stable in time. The plane normal jumps from this point to several other points and back indicating that many different planes are found.

In more detail, these plots show that there is a possibility of finding planes with a comparably high number of satellites ($\sim$12 or even higher for halo B) with reasonably low rms values ($\sim$15 kpc) in the most recent $\sim$1 Gyr. However, regarding the rms value and the number of satellites in the plane after this most recent phase, these planes are not stable. The rms value varies between 10 kpc and 18 kpc (upper panel) and between 9 kpc and 16 kpc (lower panel) for different time steps. The number of satellites varies for different time steps. For halo A (upper panel) the variation is only by about 1 satellite, and it stays nearly constant at 12 satellites for about 3 Gyr, and for halo B (lower panel) the plane gets richer in the past few time steps but then the number of satellites declines and fluctuates strongly around 12 satellites in the plane. As the right panel of the figure shows, the normal vector of halo A changes rapidly over the first Gyr (starting out from $(\phi,\theta)=(0,0)$), and afterwards changes more slowly for a short time, implying some kind of temporal coherence of the plane during this time and then around 4 to 5 Gyr in the past it varies quite a lot. For halo B the variation in the orientation of the plane is equally strong over the whole time period indicating different planes at each time step although all plane orientations seem to cluster around $(\phi,\theta)\sim(0,0)$. 

If one only follows the satellites in the plane at redshift $z=0$ back in time the behavior is even worse. For both haloes shown, the planes found at earlier cosmic times have significantly higher values of the plane thickness. For halo A the plane thickness rises nearly constantly from $\sim$ 15 kpc at redshift $z=0$ to $\sim$35 kpc at redshift $z=0.5$, $\sim$ 5 Gyr in the past. And for halo B the plane thickness rises from $\sim$ 13 kpc at redshift $z=0$ up to $\sim$ 40 kpc at redshift $\sim$ 0.4 - 0.5. At the same time the plane normal shows similar behavior as before. There seems to be no strong temporal coherence of the plane orientation and only slight clustering behavior of the plane orientations. This shows, in contrast to the previous plots where the plane thickness stays relatively constant over time with the number of satellites in the plane going down, that when the number of satellites is fixed, the planes puff up their thickness fast.
This leads to the conclusion that planes of satellites are only thin at present time and have been significantly thicker in the past. Furthermore the fluctuation of the plane orientation shows, that there seems to be no uniquely defined plane over a longer cosmic time period, leading to the impression that these planes are not stable.
Thus, although we are able to find rich and thin planes with apparent co-rotation, our analysis indicates that they do not seem to be kinematically coherent in the sense, that they are stable over time. One distinct plane appears to be a short lived system containing a significant amount of chance aligned satellites. 

\section{Discussion}
\label{sec:discussion}

We have run 21 high-resolution, fully cosmological ``zoom-in" Dark Matter only simulations of Andromeda-mass haloes to investigate planes of satellites resembling the one observed around our neighboring galaxy Andromeda. The sample of host haloes for this paper is constructed such that we select main haloes with different concentrations aiming at haloes with high, average and low concentration compared to the mean relation found by \citet{Dutton2014}. From these haloes we select the 30 most massive sub-haloes at infall time as the ones most likely to host the luminous satellites. As shown in Figure \ref{fig:rad_dist}, the radial distribution of the selected sub-haloes is similar to the observed radial distribution. Planes of satellites among these samples are then selected by a simple spatial selection criteria as the richest planes in number of members and if there are two planes of equal member numbers we select the thinner one. For simplicity and because we do not have local group analogues, we select satellites from a spherical volume rather than modeling a non-spherical volume such as the PAndAS footprint. This seems to be no major issue since \cite{Gillet2015} showed that planes of satellites found in spherical and PAndAS like volumes reveal comparable properties. After specifying the best-fitting plane by this criteria, the kinematics of these planes is investigated. 

We like to mention, that the simulations performed for this work are dark matter only ones, which do not capture the influence of baryonic physics. Therefore, the selection of the luminous satellites according to their mass at the infall time is one questionable point that could be alleviated by using hydrodynamical simulations which is beyond the scope of this paper and will be left for future work. However, the main influence of baryonic physics on our analysis is expected to be in the robust selection of luminous satellites as well as the influence of the galactic disc on the satellites in the inner region of the halo. Additionally hydrodynamical simulations will enable us to investigate the orientation of the plane of satellites compared to the disc of the galaxy. \par
Nevertheless, comparison to the work done by \citet{Gillet2015} shows that our planes compare well to the ones found in hydrodynamical simulations. Unfortunately, one draw back of hydrodynamical simulations is, up to now it is not possible to run a statistical sample of galaxies in high-resolution. E.g. \cite{Gillet2015} have only two local group analogues for their analysis, which means just one high-resolution simulation. Furthermore, they selected their galaxies to reside in a special environment, with a second massive galaxy close by. For this work we did not restrict our galaxies to have a companion, neither did we exclude such galaxies. Still our results are comparable to theirs which leaves the impression that a local group environment seems to play no major role (see also \cite{Pawlowski2014a}).\par
Another interesting point would be to investigate the effect of host halo mass on the occurrence of planes of satellites since previous studies as well as this work focussed only on Andromeda-mass like halos. Furthermore we focussed on a sample size of 30 satellites to mimic the case of Andromeda. In general it would be interesting to also investigate sample sizes of $\sim$ 50-100 sub-haloes to see wether planar structures are still visible at larger sample sizes since recently ever more satellites around MW and Andromeda are discovered. This is however difficult since it is unclear which of the smaller satellites would actually form stars, if at all, or, if they form stars, which are then observable. These questions can be probably answered in future work.

\subsection{Are the obtained planes unique?}
When selecting planes there is some ambiguity in defining the ``best" plane. As Fig. \ref{fig:2d_significance} shows there is the possibility to select planes richer in number but with a larger thickness or planes less rich in number but significantly thinner. A priori there is no natural or obvious metric to select the ``best" plane. This makes it hard to even define what a plane of satellites should look like to be regarded as a plane. Therefore, we decided to select always the plane richest in number and if there are two such planes consisting of the same number of satellites we select the thinner one. Recently this ambiguity of defining the best plane was investigated by \citet{Cautun2015}. They also find that planes of satellites show a large diversity in their properties comparable to our findings in Fig. \ref{fig:2d_significance}. They conclude that planes of satellites are common in \lcdm ($\sim$ 10\% of the haloes host planes) in agreement with our findings.

\subsection{Are our planes of satellites thin, extended and rich in number?}
All our satellite planes show a thickness smaller than $\sim$17 kpc and are more extended than $\sim$120 kpc (see Fig. \ref{fig:rms_per_par}) while at the same time including a reasonable amount of satellites (12-16). The division of haloes into high, average and low concentration haloes demonstrates that there is a clear dependence of the plane thickness on halo concentration (see also Paper I). A Kolmogorov-Smirnov test on the radial distribution of satellites of different concentration samples shows that the satellite distributions of the different concentration haloes are well comparable and high concentration haloes do not show a radially more concentrated distribution (see also Appendix A for more tests).
As the left panel of Fig. \ref{fig:rms_per_par} shows, the thinnest planes are only found in high concentration haloes, having the right thickness of $\sim$$15$ kpc and also showing the right radial extension to be comparable to the plane around Andromeda. However, comparing the three dimensional radial extension of planes is ambiguous since it can be biased by single satellites with a very high value of radius. The actual observable parameter is the two-dimensional extension of the plane on the sky. 
As we show in the right panel of Figure \ref{fig:rms_per_par} all of our planes are comparable to the value of Andromeda. Furthermore, only high concentration haloes are able to come close to the observed plane, consisting of 15 satellites in a plane (compare Fig. \ref{fig:abs_num}) of thickness $\sim$15 kpc with 13 co-rotating satellites. The reason why high concentration haloes are able to produce thinner and richer planes compared to average and low concentration haloes is their formation topology. High concentration haloes form at the nodes of the cosmic web and accrete matter along thin filaments, while average and low concentration haloes form later and accrete their matter less anisotropically. Thereby it is not the infall time of the satellites which makes the plane thinner but rather the infall direction along filaments.

Recently \cite{Pawlowski2015} published new results for the Milky Way plane of satellites including more, newly discovered satellites. The new values for the plane are: $N_{\rm in}=27$ and $\Delta_{\rm rms}=29.3$ kpc or including new satellite objects these values are $N_{\rm in}=38$ and $\Delta_{\rm rms}=30.9$ kpc . These are nearly twice (three times) as much satellites in the plane than we find for our halo. However, the rms value is also significantly higher, but this is no major problem. Looking at Figure 2 of Paper I we see, that it is easy to obtain planes with a thickness of $\sim30$ kpc which include about 20 satellites when considering a sample of 30 satellites. Raising the sample size up to e.g. $\sim$ 40 satellites to account for the new satellite objects in the Milky Way, we expect to find a comparable number of satellites in the plane with a comparable thickness. This is also found by \cite{Gillet2015}, which included sample sizes of 50 and 100 satellites in their analysis and found planes consisting of $\sim$ 30 satellites with a rms thickness of about 15 kpc. 

\subsection{Is the sign of line-of-sight velocity a stable measure for co-rotation?}
The kinematical information of the observed satellites around Andromeda and for most of the satellites of the Milky Way is only 1-dimensional and consists of the line-of-sight velocity. The co-rotation of the plane is inferred by this information, lacking full 3-dimensional information. In this work we use the advantage of the simulations with access to full 3-dimensional velocity information to investigate in detail the co-rotation of planes of satellites within the frame of \lcdm. If we infer the co-rotation of satellites in the same way as in the observations by using the line-of-sight velocity we find quite a good agreement between our planes and the observed ones (see e.g. Fig. \ref{fig:planes} for a visual impression).

When counting the number of co-rotating satellites via the line-of-sight velocity our satellite samples show a quite high co-rotation fraction up to 100\% for planes consisting of $\sim$10 satellites. As Fig. \ref{fig:corot_grid} shows, with 15 satellites in the plane and 13 co-rotating ones and a thickness smaller than 15 kpc, our high concentration haloes are able to resemble the plane of satellites observed around Andromeda.
However, the counting of the number of co-rotating satellites is done for one specific viewing angle and changes when looking at the edge-on plane from a different direction. As we show in Figure \ref{fig:rot_frac_angle} the line-of-sight co-rotation counting for different viewing angles shows a very high scatter of $\sim$2-3 satellites or $\sim$30\%. Very high co-rotation fractions are only reached for special viewing angles. Furthermore a comparison to a sample with randomized velocity vectors shows that the line-of-sight counting of this sample is fully comparable to our simulations.
Thus we conclude that the inference of co-rotation via the line-of-sight velocity is not a robust measure. This is in good agreement with the work of \citet{Phillips2015} in which they show that the inferred co-rotation of satellite planes in the Sloan Digital Sky Survey \citep[SDSS]{York2000} by opposite satellite pairs \citep{Ibata2014} is comparable with random noise when carefully investigated (however, see also \cite{Ibata2015}).

Satellite planes in the simulations which show a high degree of kinematical coherence measured in the 1-dimensional case using line-of-sight velocity information turn out to have a co-rotation fraction at most as high as 70\% when using more robust measures like an analysis of the angular momentum vectors (compare Fig. \ref{fig:ang_mom_obs}). We thus conclude from the analysis of our simulations that a more reliable measure of the co-rotation is given by the analysis of the angular momentum vectors like done for the Milky Way satellites, although it will be difficult to perform for Andromeda satellites.
 
\subsection{What does the kinematics of the satellites in the plane look like?}
The plane of satellites is just one snapshot in time. Disregarding the full three-dimensional velocity information can lead to incorrectly assigning satellites to the plane which only fly by. To quantify how many such interlopers or chance alignments of satellites occur, we come up with a simple measure of plane membership.
By measuring the velocity of the satellites in the plane in terms of the total velocity and comparing it to the case of an isotropic velocity distribution we get a handle on the number of satellites fully orbiting in the plane. From this analysis we conclude that for the simulations about 30\% of the satellites are just chance aligned with the plane (see Fig. \ref{fig:real_planes}). This can also be concluded, when comparing the direction of the angular momentum vectors of the satellites with the plane normal done in Fig. \ref{fig:hist}. Satellites orbiting in a plane should show angular momentum vectors pointing in the same direction or opposite direction as the plane normal. 
Most planes show indeed a slight clustering of their angular momentum vectors around the direction of the plane normal but with quite a high scatter consistent with about 30\% of interlopers. However, the planes found in our simulation compare well with the findings of \citet{Pawlowski2013} for the Milky Way or show even a smaller scatter around the plane normal although the planes of this work are selected by a completely different, spatial selection criteria. A comparison to a sample with random velocities shows that the angular momentum analysis is better suited to measure the co-rotation fraction of satellites in the plane compared to the line-of-sight method. 

\subsection{How does the plane evolve over time?}
The fact that the planes in our simulations contain a significant amount of interlopers leads to the question how this influences the stability of the planes over time. Planes in low concentration haloes thicken up by about a factor of 4 while in high and average concentration haloes the planes only thicken up by a factor of $\sim$2. A detailed look at the movement of satellites in the planes of halo A and B over the last 5 Gyr indicates that the plane is either not a stable system (halo A) and just at present day resembles a thin extended plane or that the plane consists of a stable backbone including about 70\% of the satellites in the plane (halo B) and a fraction of 30\% of interlopers aligning at present time with the plane. This can be concluded from Figure \ref{fig:orbit}, where the orbits of the satellites in the plane for the last 1.2 Gyr are shown. However, this analysis includes only the satellites found to be in a plane at present day and does not account for plane precession. Applying the plane finding algorithm to the sample of the 30 most massive satellites at infall time at different time steps in the simulation shows, that at every time step one is able to find a different plane instead of a slightly precessing one. This is shown in the right panel of Figure \ref{fig:thickening_time} where the orientation of the plane normal at different time steps is plotted. Accordingly, the left panel of Figure \ref{fig:thickening_time} shows, that the root-mean-square thickness is varying much while the number of satellites in the plane is going down. Looking for planes with a fixed number of satellites in the plane at different time steps shows the same result. Going back in time the plane thickens up very fast and the orientation of the plane is changing a lot.
Thus the planes of satellites in our simulations, thin and extended at present time, appear not to be a stable systems. At most, the planes arise out of a backbone of coherently rotating satellites plus a significant amount of chance aligned satellites. This may indicate that the plane of satellites around Andromeda might share the same properties as the planes found in our simulations, and thus might not be a stable system.

\section{Conclusion}
\label{sec:conclusion}

In this paper we investigated the occurrence of planes of satellites in 21 high-resolution (10 million particles) dark matter only cosmological ``zoom-in" simulations of Andromeda-mass dark matter haloes. We select dark matter haloes for re-simulation according to their formation times rather than selecting an unbiased sample. The planes found show similar properties as the one observed around our neighboring galaxy Andromeda (M31) regarding their spatial and kinematical properties, where the kinematic information is based on line-of-sight velocity measurements. Additionally, we do an angular momentum analysis of satellites in the plane using the full 3-dimensional velocity information similar to what \cite{Pawlowski2013} have done for the 11 classical Milky Way satellites. We are able to find planes of satellites simultaneously showing similar properties to the plane around Andromeda and Milky Way. Our results are summarized as follows:

\begin{enumerate}
\item Our satellite samples of the 30 most massive satellites at infall time show similar spatial distributions of satellites as the one observed around Andromeda, as Figure \ref{fig:rad_dist} shows.

\item Using a simple plane finding algorithm, we obtain planes with a variety of different values for the characteristic values of number of satellites in the plane ($N_{\rm in}$), their thickness ($\Delta_{\rm rms}$), their extent ($\Delta_{\parallel}$) and the number of co-rotating satellites ($N_{\rm corot}$) (compare Fig. \ref{fig:2d_significance}) in agreement with recent studies by \citet{Cautun2015}. From the variety of planes we select the ones richest in number and with the smallest thickness.

\item When selecting the richest and thinnest planes, our simulations reproduce the properties like number of satellites in the plane, the thickness and the extent of the observed plane well (see Fig. \ref{fig:abs_num} and left panel of Fig. \ref{fig:rms_per_par}). But caution has to be taken when comparing the simulations to the observations. The very large extent of the observed plane is biased high by $\sim$ 40 kpc due to one satellite outside the virial radius. In our simulations we are not considering such satellites since we only select satellites within the virial radius. Therefore a better comparison is the projected extent of the plane on the sky (right panel of Fig. \ref{fig:rms_per_par}) for which our planes are fully comparable to the observed plane. Furthermore, the results obtained in this paper, particularly those for the number of satellites in a plane and the number of interlopers, agree well with the results of other studies, like e.g. \cite{Gillet2015}, who used hydrodynamical simulations of two local group analogues. 

\item We further find that earlier forming haloes (high concentration) show thinner and slightly richer planes compared to average and late forming haloes (average, low concentration haloes) while there is no dependence of concentration on the extent of the planes. 

\item Including kinematical information obtained via the line-of-sight velocity of the satellites in the plane, we find planes similar to the observed one ($N_{\rm in}=15$, $N_{\rm corot}=13$, $\Delta_{\rm rms}= 12.6 \pm 0.6$ kpc). Our planes do not only compare well in the number of satellites in the plane ($N_{\rm in}$), their thickness ($\Delta_{\rm rms}$) and their extent ($\Delta_{\parallel}$) with the observations but also in the number of co-rotating satellites ($N_{\rm corot}$) obtained via the line-of-sight velocity information (Fig. \ref{fig:corot_grid}). However, the planes in the simulations are not kinematically coherent. Therefore, from our \lcdm simulations we predict a high perpendicular velocity ($\sim$100 km/s) for some of the satellites in the observed plane, which will be measurable as proper motion of the satellites perpendicular to the plane in future surveys. 

\item Investigation of the dependence of the line-of-sight velocity count of co-rotating satellites on the specific viewing angle onto the edge-on plane shows that this is not a robust measure of co-rotation (Fig. \ref{fig:rot_frac_angle}). This method uses just one component of the 3-dimensional velocity to infer the co-rotation and thus appears to be erroneous. An analysis of the angular momentum vectors shows (Fig. \ref{fig:hist}) that this latter method gives much better results on the inferred co-rotation fraction. 

\item Combination of the angular momentum vector analysis and an additional investigation of the planes for interlopers reveals a significant fraction of about 30\% chance aligned satellites (Fig. \ref{fig:real_planes}).

\item Tracking the satellites back in time and investigating earlier time steps of the simulations for planes of satellites shows that the planes in our simulations are not a long-lived system (Fig. \ref{fig:thickening_time}). They thicken up on short time scales and are hardly detectable after few hundred Myr. There is a strong dependence on halo concentration with the lower concentration haloes thicken up strongest. Investigation of the orbits of satellites in the plane (Fig. \ref{fig:orbit}) shows that a significant fraction of the satellites in the simulations does not rotate in the plane. 
\end{enumerate}

Thus we conclude that our simulations of Andromeda mass haloes in the framework of \lcdm are able to reproduce planes of satellites similar to the one around Andromeda and around Milky Way. However, the planes in our simulations in their present configuration turn out to be a rather transient occurrence with about 30\% of chance aligned satellites. Which may indicate that also the observed planes may not be kinematically coherent structures. This prediction for satellite planes from \lcdm could be tested by measuring the proper motion of the satellites in the plane.

\section*{Acknowledgments}

The authors acknowledge support from the
   Sonderforschungsbereich SFB 881 “The Milky Way System” (subproject
   A2) of the German Research Foundation (DFG).  Simulations have been
   performed on the {\sc theo} cluster of the Max-Planck-Institut fuer
   Astronomie at the Rechenzentrum in Garching. Part of the analysis for this work made use of the {\sc{pynbody}} package \citet{pynbody}.
   TB likes to thank Marcel Pawlowski for helpful discussions.



\bibliography{astro-ph.bib}

\appendix
\renewcommand{\thefigure}{A\arabic{figure}}

\setcounter{figure}{0}
\vspace*{-.5cm}
\section*{Appendix A: Correlation of minimal thickness of planes with different halo parameters}

\begin{figure}
\centering
\hspace*{-1.1cm}\includegraphics[width=.4\textwidth]{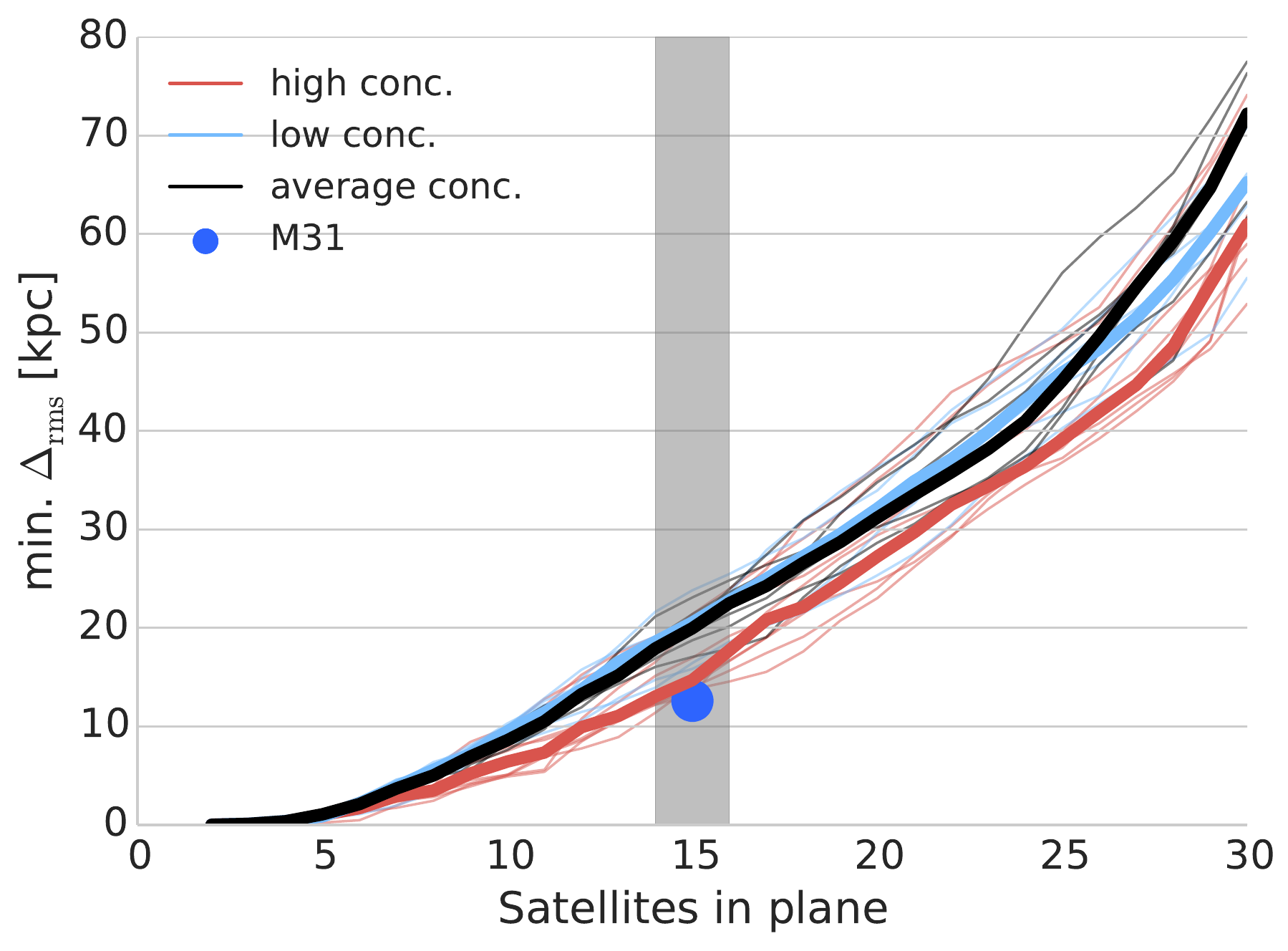}
\includegraphics[width=.49\textwidth]{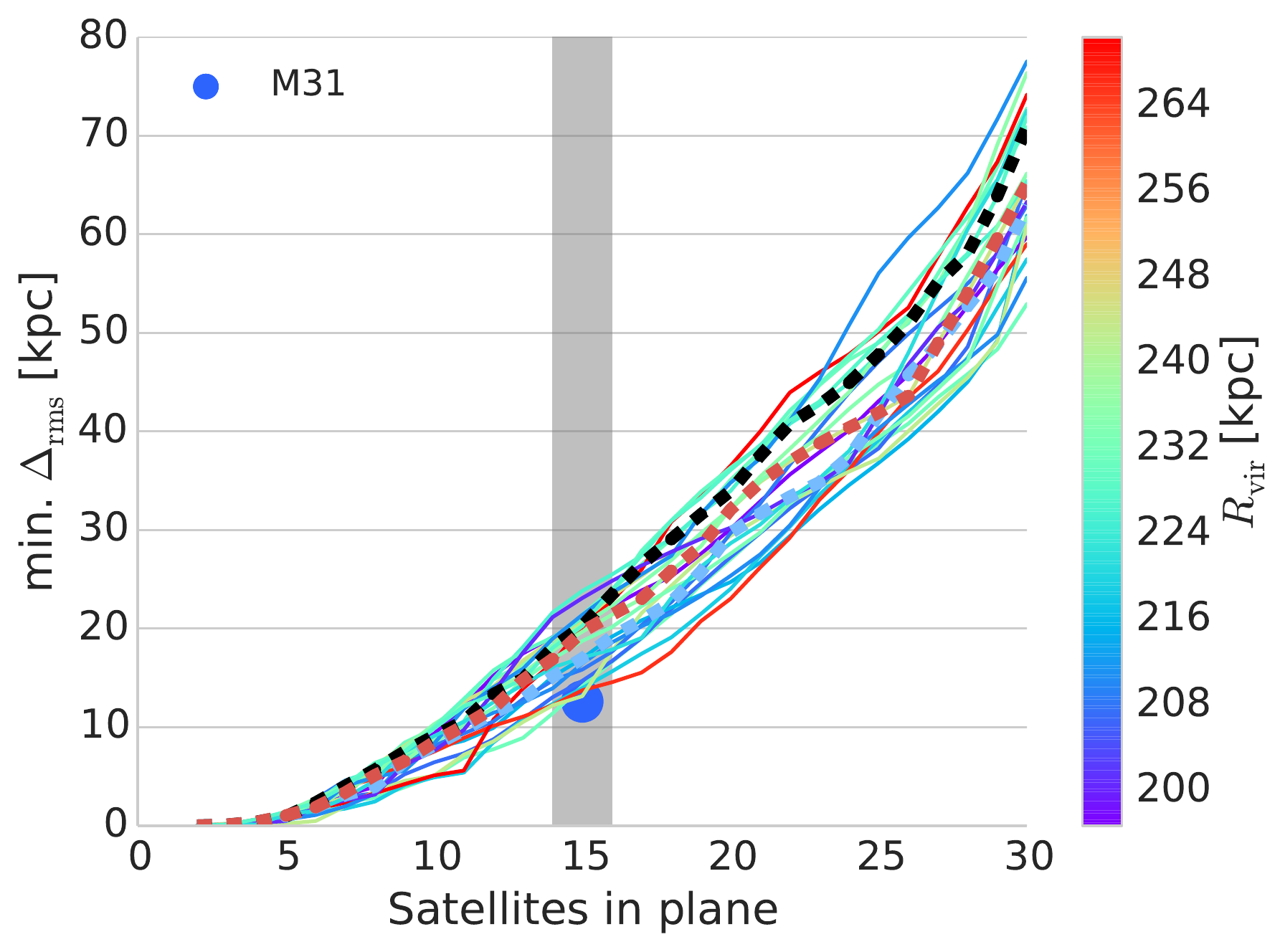}
\includegraphics[width=.49\textwidth]{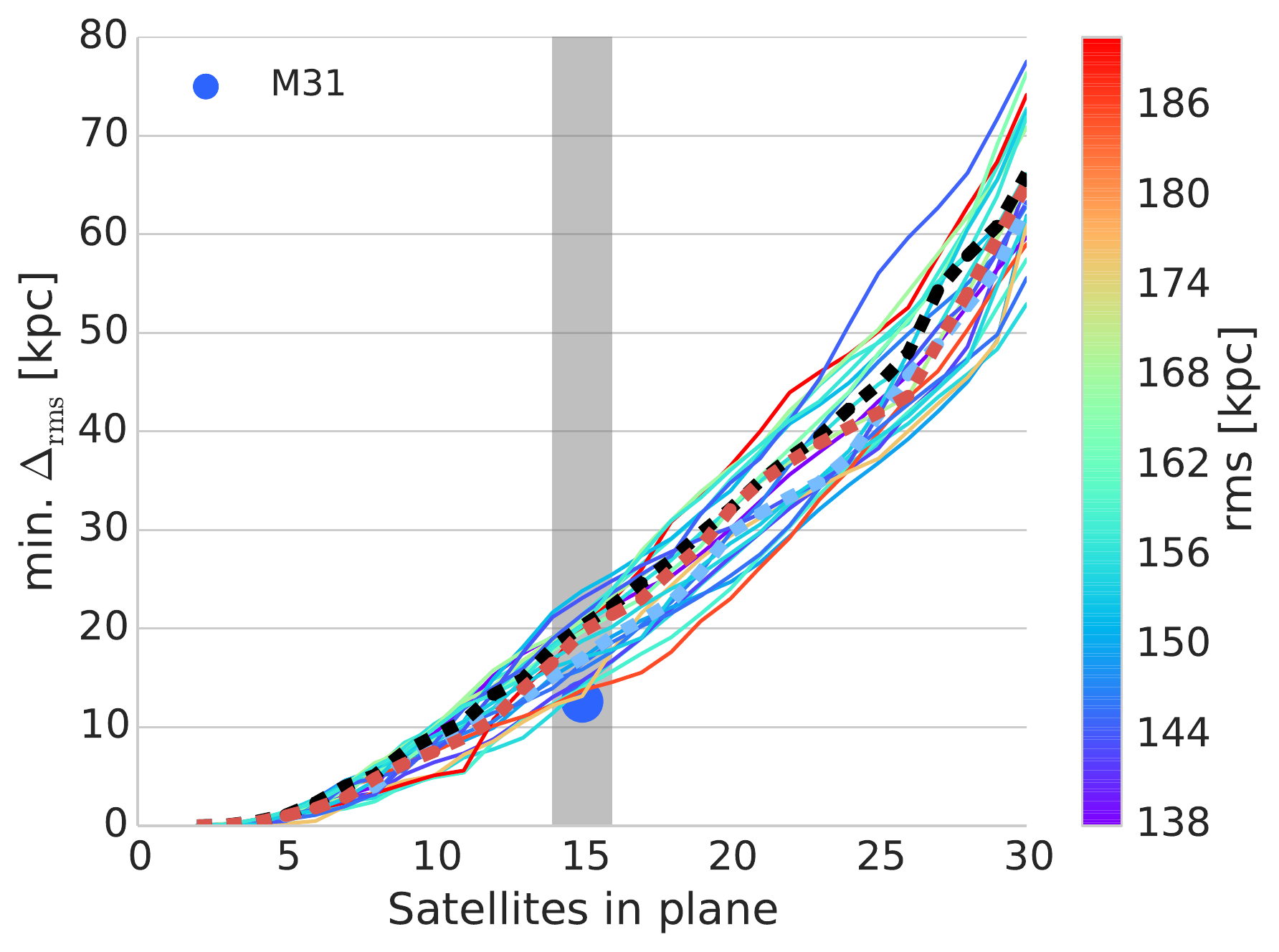}
\vspace*{-.5cm}
\caption{Minimal rms thickness, min. $\Delta_{\rm rms}$, of planes as a function of number of satellites in the plane. Each thin line represents a different dark matter halo. \emph{Upper panel}: Color coding shows the division in high (red lines), average (black lines) and low (blue lines) concentration haloes. The thick lines show the corresponding median values of each sub-sample. \emph{Middle panel}: Lines are color coded by the virial radius of the host halo. Dashed lines show the median values for sub-samples consisting of the seven haloes with highest values of virial radius (red dashed line), with average (black dashed line) and lowest values of virial radii (blue dashed line). \emph{Lower panel}: Color coding shows the root-mean-square value of the satellite's radii as a measure of the compactness of the satellite distribution. Again thick dashed lines show the median values for the seven highest (red dashed line), average (black dashed line) and lowest (blue dashed line) root-mean-square values.}
\label{fig:corr}
\end{figure}

In this section we compare the correlation of plane thickness with concentration of the halo to two other possible correlations. In Figure \ref{fig:corr} we show the correlation of plane thickness with concentration (upper panel), with virial radius (middle panel) and with the root-mean-square value of the radius of the satellite samples (bottom panel). There is a clear correlation of finding thinner planes in high concentration haloes. While the median trend for the correlation with virial radius and rms radius of the satellite distribution shows that thinner planes seem to be found in haloes with smaller virial radius or smaller rms values of the satellite distribution. However, we like to stress that nevertheless the thinnest planes are found in the haloes with the highest virial radii/rms values. This makes us conclude that the correlation with concentration is real and not due to a more concentrated satellite distribution or smaller virial radii in high concentration haloes.

\bsp	
\label{lastpage}
\end{document}